\documentclass[10pt]{article}

\usepackage{amsmath,amssymb,amsthm}
 
\def \vs {\vskip 0.2cm}

\def \n {\noindent}

\def \E {{\bf  \, E  }}

\def \P {{\bf P}}

\def \Tr {{\mathrm{\, Tr}}}

\def \bB {{\mathbb B}}

\def \bE {{\mathbb E}}

\def \bH {{\mathbb H}}

\def \bI {{\mathbb I}}

\def \bN {{\mathbb N}}

\def \bT {{\mathbb T}}

\def \bU {{\mathbb U}}

\def \bX {{\mathbb X}}

\def \bV {{\mathbb V}}

\def \bW {{\mathbb W}}

\def \bY {{\mathbb Y}}

\def \CZ {{\cal Z}}

\def \CA {{\cal A}}
\def \CB{{\cal B}}

\def \CE {{\cal E}}
\def \CC {{\cal C}}

\def \CG {{\cal G}}
\def \CH {{\cal H}}

\def \CN {{\cal N}}
\def \CT {{\cal T}}

\def \CM{{\cal M}}
\def \CW {{\cal W}}
\def \CD {{\cal D}}
\def \CI {{\cal I}}
\def\CL {{\cal L}}

\def \CP {{\cal P}}
\def \CR {{\cal R}}

\def \CO {{\cal O}}

\def \CS {{\cal S}}

\def \CX {{\cal X}}

\def \CZ {{\cal Z}}

\def \D {\Delta}

\def \G{{\Gamma}}

\def \a {\alpha}

\def \b {\beta}
\def \g {\gamma}
\def \l {\lambda}
\def \r {\rho}

\def \vk {\varkappa}

\def \k {\kappa}
\def \vr {\varrho}

\def \t {\tau}

\def \d {\delta}
\def \U {\Upsilon}
\def \u {\upsilon}
\def \th {\theta}

\def \vep {\varepsilon}

\def \vp {\varphi}

\def \L {\Lambda}

\def \la {\langle}
\def \ra {\rangle}

\def \bb {{\breve \beta}}

\def \brW {{\breve \CW}}

\def \bt {{\breve t}}

\def \fb {{\mathfrak b}}

\def \fc {{\mathfrak c}}

\def \fA {{\mathfrak A}}

\def \fB {{\mathfrak B}}

\def \fR {{\mathfrak R}}

\def \fs {{\frak s}}

\def \fS {{\frak S}}

\def \fa {{\mathfrak a} }

\def \fm {{\mathfrak m}}

\def \RT {{\mathrm T}}

\def \rt {{\mathrm t}}

\def \rI {{\mathrm I}}

\begin{document}
\title{ On High Moments 
of Strongly  Diluted Large Wigner Random Matrices 
}

\author{O. Khorunzhiy\\ Universit\'e de Versailles - Saint-Quentin \\45, Avenue des Etats-Unis, 78035 Versailles, FRANCE\\
{\it e-mail:} oleksiy.khorunzhiy@uvsq.fr}
\maketitle
\begin{abstract}
 We consider a dilute version of the Wigner ensemble of $n\times n$ random real symmetric  \mbox{matrices}
$H^{(n,\rho )}$, where  $\rho$ denotes the average number of non-zero  {elements}
per row.
We study  the
asymptotic properties of the moments  
$M_{2s}^{(n,\r)}= \E \Tr (H^{(n,\r)})^{2s}$  in the limit
 when  $n$, $s$  and $\r$ tend to infinity.
 
Our main result is that  the sequence 
$M_{2s_n}^{(n,\r_n)}$ with $s_n = \lfloor \chi \r_n\rfloor $,  $\chi >0$, $\r_n\to\infty $ and $\r_n = o( n^{1/5})$ is asymptotically close
to a sequence of numbers $n \hat m_{s_n}^{(\r_n)}$, where $\{\hat m_s^{(\r)}\}_{s\ge 0}$ are   determined 
by an explicit recurrence that involves the second and the fourth moments of the random variables
 $(H^{(n,\rho )})_{ij}$, $V_2$ and $V_4$, respectively.
This recurrent relation generalizes  the one that  determines  the moments of the Wigner's semicircle
law
given by $ m_s= \lim_{\r\to\infty} \hat m_s(\r)$, $s\in \bN$. It shows that the spectral properties of random matrices
at the edge of the limiting spectrum regarded in the asymptotic regime of the strong dilution essentially differ from
 those observed in the cases of moderate and weak dilution, where the dependence on the fourth 
moment $V_4$ does not intervene.

\end{abstract}
\vskip 0.3cm

{\it Running title:}  Strongly  Diluted Wigner Random Matrices


\section{Introduction, main results and discussion}

Spectral theory  of high dimensional random matrices
represents  an intensively developing branch of modern  mathematical physics
that reveals deep links between probability theory, analysis, combinatorics 
 and other various fields of mathematics (see monographs \cite{AGZ, M}). 
The first studies of  spectral properties  of random matrices of infinitely increasing dimensions were
started by E. Wigner (see e. g. \cite{W}), 
where the ensemble of real symmetric matrices of the form 
$$
(A^{(n)})_{ij} = {1\over \sqrt n} a_{ij}
\eqno (1.1)
$$
was introduced and the
limiting eigenvalue distribution of $A^{(n)}$, $n\to\infty$ was determined explicitly.
 The random matrix entries of $A^{(n)}$ (1.1)
are given by jointly independent random variables $\{a_{ij}, i\le j\}$
that have all moments finite and the odd moments zero. At present, this ensemble is referred  
as the Wigner ensemble of random matrices.  It was proved in \cite{W} that   the eigenvalue distribution 
of $A^{(n)}$ converges in average as $n\to\infty$  to a non-random limit with the density of the semi-circle form.
At present this  convergence  is widely known  
as the semicircle (or Wigner) law for random matrix ensembles. 

The semicircle law was generalized  in several  directions. One group of 
generalizations concerns the properties  of the  probability distributions of elements $a_{ij}$, 
another one is related with the studies of the spectral norm of $A^{(n)}$
and other local properties of the eigenvalue distribution at the border of the limiting spectrum 
or 
inside of it. 

 A large number of works is related with   various generalizations  
 of the Wigner ensemble that involve  modifications of the random matrix entries.
In the present paper we study 
one of such  generalizations   given by the
ensemble of dilute random matrices. 
We consider  a family of real symmetric
random matrices $\{H^{(n,\r)}\}$ whose elements are determined by equality
$$
\left( H^{(n,\r)}\right)_{ij} = a_{ij}\, b_{ij}^{(n,\r)}, \quad 1\le i\le j \le n,
\eqno (1.2)
$$
where $\fA= \{a_{ij}, 1\le i\le j\}$ is an infinite family
of jointly independent identically distributed random variables and
  $\fB_n= \{ b^{(n,\r)}_{ij}, 1\le i\le j  \le n\}$
is a family of jointly independent between themselves random variables that are also independent
from $\fA$.
We denote by \mbox{$\E = \E_n$} the mathematical expectation with respect to the measure
$\P= \P_n$ generated by random variables 
$\{\fA,\fB_n\}$.
We assume that the probability distribution of random variables $a_{ij}$
 is symmetric and denote their even moments by
$$V_{2l}= \E (a_{ij})^{2l}, \quad l=1,2,3,\dots$$

Random variables $b_{ij}^{(n,\r)}$ are proportional to the Bernoulli ones,
$$
b_{ij}^{(n,\r)}= {1\over \sqrt \r}
 \begin{cases}
  1-\delta_{ij} , & \text{with probability $\r/n$}, \\
0, & \text {with probability $1-\r/n$,}
\end{cases}
\eqno (1.3)
$$
where $\d_{ij}= \d_{i,j} $ is the Kronecker $\d$-symbol.
In the case when the dilution parameter $\r$ is equal to $n$,  one gets  the Wigner ensemble
of real symmetric random matrices $A_n$ (1.1). 
Let us note that the random matrix $B^{(n,\r)}$ with the entries $\sqrt \r \,b_{ij}$ (1.3) can be regarded as 
the adjacency matrix of the Erd\H os-R\'enyi random graph \cite{BB}. In this interpretation, the dilution parameter $\r$
represents the average degree of a given vertex of  the graph.

\vskip 0.2cm
The initial   interest in the dilute versions of Wigner ensemble  was motivated by   
 theoretical physics studies (see for instance, the pioneering works \cite{MF}, \cite{RB} and the review \cite{KKPS} for more references), 
 where the  
spectral properties of large systems with a number of  broken interactions were  considered.
This kind of random matrices is also important  in the studies of various mathematical models,  
such as
random graphs \cite{E,FK,KS}  and  many others.

\vskip 0.2cm

In the present paper  we study the asymptotic behavior of  the moments of $H^{(n,\r)}$ given by expression 
$$
M_{2s}^{(n,\r) } = \E \left(\sum_{i=1}^n  (H^{(n,\r)})^{2s}_{ii}\right)= \E \left(\Tr\,  (H^{(n,\r)})^{2s}\right).
$$
The moment method represents an effective tool of  the spectral theory.
It is used in the studies of the spectral properties of  large random matrices since
the
pioneering works of E. Wigner \cite{W}. In particular, the semicircle law was proved initially 
by the convergence  of the moments $M_{2s}^{(n,n)}$ in the limit of infinite $n$ and given $s$. 
The principal idea of the Wigner's approach is to consider the trace of the product of random matrices
as the sum over the family of trajectories of $2s$ steps and then to compute the weights of these trajectories
given by the mathematical expectation of the products of corresponding random variables.

The  moments $M_{2s}^{(n,n)}$ of Wigner random matrices $A^{(n)}$ (1.1)  
in the limit $n\to\infty$ with infinitely increasing 
$s = s_n$  were studied in a long series of papers, where the  eigenvalue distribution
at the edge of the limiting spectra was studied in more and more details
\cite{BY,FK,G}.  The crucial step has been performed in papers
 \cite{SS1,SS2}, where the original Wigner\rq{}s moment method has got powerful and deep improvement.
In these studies, 
 the Tracy-Widom law for random matrices $A^{(n)}$ 
established in the case of  normally distributed entries  $a_{ij}$
is shown to be true in the general case of arbitrary probability distribution of $a_{ij}$ \cite{S}. 
This result is obtained by analysis of the high moments 
$M_{2s_n}^{(n,n) }$ in the limit $n, s_n\to\infty$ with $s_n = O(n^{2/3})$.

\vs 
The high moments of large dilute random matrices $H^{(n,\rho_n)}$ (1.2)  were studied in \cite{K3} in the asymptotic regime
when $\r= \r_n = O(n^{\a})$ with $2/3 < \a<1$. It was proved that the limiting expression of the moments  
$M_{2s_n}^{(n,\r_n)}$ with $s_n = O(n^{2/3})$ coincides with that of 
 the moments of the Wigner random matrices $M_{2s_n}^{(n,n) }$.
 This fact  can be regarded as an evidence 
of the universal behavior of the local eigenvalue statistics for weakly dilute random matrices,
i.e. when the dilution parameter $\r$ is sufficiently large. 
In the present paper we study the opposite asymptotic regime
 of strongly dilute random matrices, i. e. when the dilution parameter 
$\r_n$ tends to infinity as $n\to\infty$ but with much lower range than before, $\r_n = O(n^{\a})$ with $\a<1/5$. 
We show that in the limit 
$$
n, \r_n\to \infty,  \  \r_n =o(n^{1/5}), \ s_n = \lfloor \chi \r_n\rfloor,\quad \chi>0,
\eqno (1.4)
$$
where $\lfloor x\rfloor$ is the integer part of $x$,
 the
limiting expressions of $M_{2s_n}^{(n,\r_n) }$ are different from  those obtained for the  Wigner random matrix ensemble.
This difference is due to the fact that the leading contribution to the moments $M_{2s}^{(n,\r)}$ in the asymptotic regime (1.4)
is given by the trajectories that generalize in certain sense the Catalan numbers  that describe the moments
of the Wigner ensemble. Up to our knowledge, these trajectories of the new type were not considered before. 
Their combinatorial properties are of their own interest
and this fact  has
 motivated the  work presented. In our studies, in particular, we obtain a number of explicit relations
 that were not known in the context of random matrices and plane rooted trees (see, for example, relation (1.9) below and 
 formulas (5.3) and (5.4)
 of Section 5).

To make more compact the   formulas we use, everywhere below we refer to  the limiting transition  (1.4) as  $(n,s,\r)\to\infty$.
Our main result is given by the following statement.

\vskip 0.5cm
{\bf Theorem 1.1.}  
{\it Assume that $V_2=1$ and that for all $1\le i \le j$ the random variables $a_{ij} $ are bounded with probability 1, 
$$
\vert a_{ij}\vert \le U.
\eqno (1.5)
$$
 There exists a constant  $\chi_0 = \chi_0( U) >0$  
such that for any given 
$0< \chi< \chi_0$
the following 
upper bound holds in the limit (1.4),
$$
\limsup_{(n,s,\r)\to\infty} {1\over n \rt_s} M_{2s_n}^{(n,\r_n)}\le 4 e^{  16 \chi  V_4},
\eqno (1.6)
$$
where 
$$
\rt_s = {(2s)!\over s!\, (s+1)!}, \quad s=0,1,2,\dots
\eqno (1.7)
$$
are the Catalan numbers.
The moments $M_{2s_n}^{(n,\r_n)} $ are given by the following asymptotic relation,
$$
M_{2s_n}^{(n,\r) } = n \hat m_{s_n}^{(\r_n)} (1+ o(1)), \quad (n,s,\r)\to\infty, 
\eqno (1.8)
$$
where the sequence $\{\hat m_s^{(\r)}\}_{s\ge 0}$  is  such that its generating function
$F_\r(z) = \sum_{s\ge 0} \hat m_s^{(\r)} \, z^s$ verifies equation
$$
F_\r(z) = 1+ z \left(F_\r(z)\right)^2 + {z^2V_4\over \r}\left( {1\over 1-zF_\r(z)}\right)^4
\eqno (1.9)
$$
with the initial condition $\hat m_0^{(\r)} = 1$.
 }

\vskip 0.2cm
{\it Remarks} 

1.  We restrict the rate of $\r_n$  by $n^{1/5}$ (1.4)    not to overload the technical part of the paper.
In fact,  it follows from the proof of  Theorem 1.1 that relations (1.6) and (1.8) can be obtained 
with   (1.4) replaced by the limit when 
$\r_n \to\infty$ and $\r_n = o(n^{1/2})$ (see formula (3.30) and  the discussion that follows it). 
Moreover, one can expect that Theorem 1.1 remains valid
in the asymptotic regime when $\r_n = n^{\a}$ with $0<\a<2/3$. This asymptotic regime  is complementary to the one
studied in \cite{K3}.  However, in the present paper we are aimed mostly at the lowest rates of $\r_n$ having a 
 particular  interest in  
the asymptotic regime  when $\r_n = O(\log n), n\to\infty$. 

\vskip 0.1cm
2. In contrast to the technical restriction (1.4), it is not clear whether condition (1.5) 
can be relaxed, especially in the case of the
asymptotic regime  when $\r_n = O(\log n), n\to\infty$ and $s_n =\lfloor  \chi \r_n\rfloor$. 
However, a part of the  estimates that concerns the tree-type walks can be proved under 
considerably less restricted conditions than (1.5) (see relation (4.5) of Section 4 below). 
 
\vskip 0.1cm
3. We will show that the numbers $\hat m^{(\r)}_s$ are uniquely determined and verify the following upper bound (cf. (1.6)),
$$
{1\over  \rt_s} \hat m_s^{(\r)} \le 4  e^{4 V_4s/\r}.
\eqno (1.10)
$$
Therefore the generating function $F_\r(z)$ (1.9) exists and is bounded in absolute value  for any given $\r$. Then it follows from (1.9)
that the limiting function $f(z)= \lim_{\r\to\infty} F_\r(z)$ exists and verifies the following relation,
$$
f(z) = 1 + z(f(z))^2.
\eqno (1.11)
$$
This equation has a unique solution that  determines the generating function of the Catalan numbers (1.7),
$f(z) = \sum_{k\ge 0} \rt_k z^k$. 

\vskip 0.1cm
4. Relation (1.8) can be rewritten in a more precise form. 
We will show that there exists a constant $C_0(U)>0$ such that 
 the following relation holds
$$
\limsup_{(n,s,\r)\to\infty} {1\over n \rt_s} \left( M_{2s_n}^{(n,\r_n)} - n \hat m^{(\r_n)}_{s_n}\right) \le {CV_4\over \r} e^{16\chi V_4}
\eqno (1.12)
$$
in the limit $n,s_n,\r_n \to\infty$ such that $s_n = \lfloor \chi \r_n\rfloor$, $0<\chi\le \chi_0$ and $\r_n = o(n^{1/6})$. 
In fact, one can show that the left-hand side of relation (1.12) admits the asymptotic expansion in powers of $\r$
and that the first terms of this expansion are given by relation
$$
{1\over n\rt_s} M_{2s}^{(n,\r)} = {1\over  \rt_s} \left( \hat m_s^{(\r)} + {1\over \r} R^{(1)}_s + o( \r^{-1})\right), \quad \r\to\infty, 
\eqno (1.13)
$$
where 
$$
R^{(1)}_s = {4V_4^2\over \rho}  {(2s)!\over (s-4)! \, (s+4)!} + { V_6\over \rho}  {(2s)!\over (s-3)!\, (s+3)!} + o(\chi), \quad \chi\to 0
$$
(see the last part of Section 4).

\vskip 0.1cm
5. As one will see,  the numbers $\hat m_{s}^{(\r)}, k\ge 0$ of (1.8) can be regarded as
a generalization of the Catalan numbers $\rt_k, k\ge 0$ in the following sense. 
The Catalan number $\rt_k$ 
counts the half-plane rooted trees of $k$ edges. Regarding the chronological run over the such a tree,
we get a closed  walk of $2k$ steps such that in its graph each edge is passed two times, in  there and back directions.
Using this terminology, we can say that  $\hat m_s^{(\r)}$ represents the sum of weighs of all closed walks of $2s$ steps
such that in their graphs each edge is passed either two or four times when counted in there and back directions. Also it is shown that
in the corresponding graph the edges passed four times do not share a common vertex
(see Section 4 for  the rigorous definition of tree-type 
 $(2,4)^*$-walks). 
 
One of the consequences of this definition of numbers $\hat m_s ^{(\r)}$ is given by the following lower bound (see  formula (5.3) of Section 5 below),
$$
\hat m_s^{(\r)} \ge \rt_s + {V_4\over \r}\cdot  {(2s)!\over (s-2)! \, (s+2)!},
$$
where the last fraction represents the number of closed tree-type walks of $2s$ steps such that their graphs
contain exactly one edge passed four times. Then in the limiting transition (1.4) we get the following 
lower bound,
$$
\liminf_{(n,s,\r)\to\infty} {1\over \rt_s} \, \hat m_s^{(\r)} \ge (1+ \chi V_4).
\eqno (1.14)
$$

\vskip 0.5cm 

Let us discuss relations of the results of Theorem 1.1 with the spectral properties  of large dilute random matrices $H^{(n,\rho)}$ (1.2).
Regarding the ordered family of real eigenvalues of the matrix 
$H^{(n,\rho)}(\omega)$, $\l_1^{(n,\r)}(\omega)\le \cdots \le \l_n^{(n,\r)}(\omega)$, we denote its spectral norm by $\l_{\max}^{(n,\r)}(\omega)$,
$$
\Vert H^{(n, \r)} (\omega) \Vert = \l_{\max}^{(n,\r)}(\omega) = \max\{ \vert \l_1^{(n,\r)}(\omega)\vert, \vert \l_n^{(n,\r)}(\omega)\vert\}.
$$
The well-known analog of the Chebyshev inequality for the deviation probability 
$$
P\left( \l_{\max}^{(n,\r)} \ge 2(1+\vep)\right) \le {1\over (2(1+\vep))^{2s}} \E \left( \sum_{j=1}^n (\l_j^{(n,\r)})^{2s}\right) = 
{1\over (2(1+\vep))^{2s}} M_{2s}^{(n,\r)}
\eqno (1.15)
$$
allows us to deduce from  estimate (1.6) the following upper bound valid for all $n\ge n_0$ with some $n_0$,
$$
P\left( \l_{\max}^{(n,\r)} \ge 2(1+\vep)\right) \le 6 e^{16  V_4 \chi }\,  {n \rt_s \over (2(1+\vep))^{2s}}. 
$$
Applying  the Stirling formula to the Catalan numbers $\rt_s$ (1.7), we get inequality 
$$
P\left( \l_{\max}^{(n,\r)} \ge 2(1+\vep)\right) \le 4 e^{16  V_4 \chi }\, {n\over s^{3/2} (1+\vep)^{2s}}.
\eqno(1.16)
$$
Remembering that  $s_n = \lfloor \chi \r_n \rfloor$, 
we deduce from (1.16) that if the sequence $(\r_n)_{n\ge 1}$ is such that $\r_n/ \log n\to\infty$ as $n\to\infty$, then   
for any given $\vep>0$
$$
\sum_{n\ge n_0} P\left( \l_{\max}^{(n,\r_n)} \ge 2(1+\vep)\right) <\infty .
\eqno (1.17)
$$
 Relation (1.17) means that $P\left( \limsup_{n\to\infty} \l_{\max}^{(n,\r_n)} \le 2\right) =1$
  in this asymptotic regime. 

Taking into account the fact that the semicircle law is  valid for the eigenvalue distribution of $H^{(n,\r_n)}$ in the limit
$n,\r_n\to\infty$ \cite{KKPS}, it is not hard to conclude that  with probability 1,
$$
\lim_{n\to\infty} \Vert H^{(n,\r_n)}\Vert = 2, \quad {\r_n\over \log n} \to\infty, \ n\to\infty.
\eqno (1.18)
$$
This statement slightly improves our earlier  results \cite{K01}, where the convergence of $\l_{\max}^{(n,\r_n)}$ to $2$ has been proved in the asymptotic regime when $\r_n = (\log n)^{1+\delta}$ with given $\delta >0$. 
Let us note however, that in the present article we prove Theorem 1.1 and (1.18) under
condition (1.5) that is more restrictive than those imposed in paper \cite{K01}. 

\vs

Returning to inequalities (1.15) and (1.16), we can rewrite them in the following  form
$$
\Psi_n(x) = P\left( \l_{\max}^{(n,\r)} \ge 2\left( 1 + {x\over h_n}\right)\right) \le 
{4n\over s^{3/2}} \exp\left\{ 2\chi \left( 8V_4 - {x\r_n\over h_n}\right)\right\}.
\eqno (1.19)
$$
The right-hand side of (1.19) shows that in the limit of infinite $n$, the probability to find eigenvalues of $H^{(n,\r_n)}$ outside of the interval
$(-2(1+x/h_n), 2(1+x/h_n)$ goes to zero provided $h_n $ is much smaller than $\r_n (\ln n)^{-1}$, i.e. the length of the corresponding interval is larger than $(\log n )/ \r_n$.

Neglecting the first factor in the right-hand side of (1.19), we can observe that to obtain a non-trivial and non-zero limit 
$\Psi_n(x)\to \Psi(x)$, it is natural to consider the scaling parameter  $h_n$ of the left-hand side of (1.19) to be of the order $O(\r_n)$. 
This  reasoning could be compared with the result widely known as the Tracy-Widom distribution  for the maximal eigenvalue
of Gaussian Unitary (and Orthogonal) Ensembles of random matrices of the form (1.1), where the limiting expression of $\Psi_n(x)$
is explicitly determined with the scaling factor $h_n = n^{-2/3}$ \cite{TW}  (see also monographs \cite{AGZ, M}). 

\vs
It should be stressed that in papers \cite{S, S2}, the existence of a non-trivial non-zero limit  of the moments 
$ \lim_{n\to\infty} M_{2s_n}^{(n,n)}= \CM(\t)$, 
$s_n = \lfloor \tau n^{2/3}\rfloor$ of the Wigner random matrices $A_n$ (1.1) 
 is shown to imply the Tracy-Widom law for the maximal eigenvalue distribution in the case of arbitrarily distributed matrix entries 
 $a_{ij}$. The limiting expression $\CM(\t)$ being independent from the the particular values of the moments $V_{2l}, l\ge 2$ of $a_{ij}$,
 the result of \cite{S} means a wide universality of the Tracy-Widom law for large random matrices. 
 
 In paper \cite{K3} we have proved that the same limiting expression $\CM(\t)= \lim_{n\to\infty} M_{2s_n}^{(n,\r_n)}$, $s_n= \lfloor \t n^{2/3}\rfloor$ is valid 
  for the ensemble of dilute random matrices (1.2) in the asymptotic regime of weak dilution, $\r_n /n^{2/3} \to\infty$. 
In contrast, when the dilution becomes moderate, i.e. when $\r_n =\zeta n^{2/3}$ with given $\zeta>0$, the estimate from below 
(see \cite{K3}, Theorem 7.1)
$$
\liminf_{n\to\infty} M_{2s_n}^{(n,\r_n)} \ge {4V_4\over \zeta \sqrt{\pi \t} } e^{-e\t^3}
\eqno (1.20)
$$
shows that  the universality of the Tracy-Widom distribution 
discussed above cannot hold in this asymptotic regime. This is because the fourth moment $V_4$
enters explicitly into the lower bound (1.20) while this is not so in the case of non-diluted or weakly diluted Wigner random matrices
according to the statements of \cite{S,S2}.

\vs Returning to the case of the strong dilution when $\r_n =o(n^{1/5})$ (1.4), 
one can see that the situation becomes even worse than that of (1.20) because  
the sequence of  moments $M_{2s_n}^{(n,\r_n)}$ diverges exponentially with respect to $s_n/\r_n$, i.e.  whenever $s_n$ is of the order greater than
$\r_n$ as $n\to\infty$. This is because the expression in the right-hand side of the upper bound 
(1.10) turns out to be the lower one with the numerical constants tuned up (see a conjecture on the numbers ${\cal N}_s^{(p,2)}$
at the end of the sub-section 5.2). 
The exponential divergence of $M_{2s_n}^{(n,\r_n)}$ can be regarded as one more argument
to the conjecture that the  scaling parameter $h_n$  (1.19) at the edge of the spectrum of large strongly diluted random matrices 
has to be switched from $n^{-2/3}$ to $\r_n^{-1}$. We postpone the study of this problem to subsequent publications.

\section{Trajectories,  walks and graphical representations}

In this section we describe the main  
components of the method we develop to study the high  moments of dilute random matrices $H^{(n,\r)}$ (1.2). 
In the  pioneering works of  E. Wigner (see e.g. \cite{W}), 
it was proposed to consider  the moments $M_{2s}$  of random matrices as a weighted sum
over paths of $2s$ steps. In the case of dilute random matrices, we can write that 
$$
 M_{2s}^{(n,\r)} = 
 \sum_{i=1}^n \E   \left( H^{(n,\r )}\right)^{2s}_{ii}  = 
 \sum_{\CI_{2s} \in \bI_{2s}(n)}
 \Pi_{a,b} (\CI_{2s}) 
= \sum_{\CI_{2s} \in \bI_{2s}(n)}
 \Pi_a(\CI_{2s}) \, \Pi_b(\CI_{2s}),
\eqno (2.1)
$$
where the sequence $\CI_{2s} = (i_0,i_1,\dots, i_{2s-1},i_0), i_k\in \{1,2,\dots, n\} $ is
regarded as a closed path  of $2s$ steps $(i_{t-1},i_{t})$ with the discrete time $t\in [0,2s]$. 
We will also say that $\CI_{2s}$
is a trajectory of $2s$ steps. The set of all possible trajectories of $2s$ steps over $\{1,\dots, n\}$ is denoted by $\bI_{2s}(n)$.
The weights
$
\Pi_a(\CI_{2s}) $ and $\Pi_b(\CI_{2s})$ are  determined
as the mathematical expectations  of the products of corresponding random variables,
$$
 \Pi_a(\CI_{2s}) = \E \left( a_{i_0i_1}\cdots  a_{i_{2s-1}i_0}\right), \quad
\Pi_b(\CI_{2s}) = \E \left(b_{i_0i_1}\cdots b_{i_{2s-1}i_0}\right).
\eqno (2.2)
$$
Here and below, we omit the superscripts in $b^{(n,\r)}_{ij}$ when no confusion can arise.

 In papers \cite{SS1,SS2} a deep and powerful generalization of the E. Wigner's approach was
 proposed by Ya. Sinai and A. Soshnikov to  study  the high moments of random matrices. Somehow different point of view has been
 developed to consider the ensembles of dilute random matrices  \cite{K01,K3,KSV}. 
 The difference between these
 two approaches is dictated by the fact that  the leading contribution 
 to the moments $M_{2s}^{(n,\r)}$ of the dilute random matrices $H^{(n,\r)}$  (1.2) 
 is given by  those trajectories $\CI_{2s}$ that have a vanishing weight 
 in the case of  non-diluted Wigner random matrices $A_n$ (1.1) studied in \cite{SS1,SS2}.
 In the first part  of the present paper we use a combination of these two approaches to study the 
 terms that give the vanishing contribution to the moments of strongly diluted random matrices (see Section 3).
 On the second stage we develop a new method that allows us to prove that in the limit of infinitely increasing dimension, non-zero contribution 
to the moments $M_{2s}^{(n,\r)}$ is given by the new kind of tree-type walks such that their weight contains the factors $V_2$
and $V_4$ only (see Section 4). We refer to this kind of walks as to (2,4)-walks. To determine rigorously the corresponding classes of trajectories,
we need to describe briefly the fundamental notions of the  methods developed in papers mentioned above.

Regarding a trajectory $\CI_{2s}$ one can determine  a {\it walk} 
$$
\CW_{2s}= \CW^{(\CI_{2s})}_{2s}= \{\CW(t), t\in [0,2s]\}, \quad 
\hbox{where } \quad [0,2s] = \{ 0, 1, 2,\dots, 2s\}, 
$$ 
that we define as a sequence
of $2s+1$ symbols  (or equivalently, letters)
 from an ordered  alphabet, say
$ {\cal A} = \{\a_1, \a_2, \dots\}$. The walk $\CW^{(\CI_{2s})}_{2s}$ is
 constructed with the help of the following rules of recurrence  \cite{KSV}. 
 Given a trajectory
$\CI_{2s}$, we write that $\CI_{2s}(t) = i_t$, $t\in [0,2s]$ and consider a subset
$\bU(\CI_{2s};t) = \{ \CI_{2s}(t'), \, 0\le t'\le t\}\subseteq \{1,2,\dots,n\}$.  We denote by
$ \vert \bU(\CI_{2s};t)\vert$ its cardinality. Then 

\vskip 0.1cm
1) $\CW_{2s}(0)= \a_1$;

2) if $\CI_{2s}(t+1) \notin \bU(\CI_{2s};t)$, then $\CW_{2s}(t+1) = \a_{\vert \bU(\CI_{2s};t)\vert +1}$;

\hskip 0.44cm if there exists $t'\le t$ such that $\CI_{2s}(t+1) = \CI_{2s}(t')$, then $\CW_{2s}(t+1) = \CW_{2s}(t')$.

\vskip 0.1cm

\noindent For example,
$
\CI_{16} = (5,2,7,9,7,1,5,2,7, 9, 7, 2, 7, 2, 7, 1,5) 
$
produces the walk
$$
\CW_{16} =  (\a_1,\a_2,\a_3,\a_4,\a_3,\a_5,\a_1,\a_2,\a_3,\a_4,\a_3,\a_2,\a_3,\a_2,\a_3,\a_5,\a_1).
\eqno (2.3)
$$
We say that the pair $(\CW_{2s}(t-1), \CW_{2s}(t))$ represents the $t$-th step of the  walk $\CW_{2s}$ and that
$\a_1$ represents the {\it root} of the walk $\CW_{2s}$. 

Taking two trajectories $\CI'_{2s}$ and $ \CI''_{2s}$ such that 
 $\CW^{(\CI'_{2s})}_{2s} = \CW^{(\CI''_{2s})}_{2s} = \CW_{2s}$, we say that they are
 equivalent, $\CI'_{2s}\sim  \CI''_{2s}$. We
  denote by $\CC_{\CW_{2s}}$ the corresponding  class of equivalence.
It is clear that
$$
\vert \CC_{\CW_{2s}}\vert = n(n-1)\cdots (n-  \vert \bU(\CI_{2s};2s)\vert+1).
\eqno (2.4)
$$

Given $\CW_{2s}$, one can introduce a {\it graphical representation}  $ g(\CW_{2s}) = (\bV_g, \bE_g)$ that can be
considered as  a kind of multigraph
with the set of vertices $\bV_g = \{\a_1, \dots, \a_{\vert \bU(\CI_{2s};2s)\vert}\} $ and the set $\bE_g$ of $2s$ oriented edges
(or equivalently, arcs) labelled by $t\in \{1, \dots, 2s\}$.
To describe the properties of $g(\CW_{2s})$ in general situations, we will 
use greek letters $\a,\b,\gamma,\dots $ instead of the symbols from ${\cal A}$.
In this case, the root of the walk  will be  denoted by  $\varrho$.
In what follows, we refer to  $g(\CW_{2s})$  simply as to  the {\it graph} of the walk  $\CW_{2s}$.
If $\CW_{2s}(t)=\gamma$, we say that $\gamma$ is the {\it value of the walk } $\CW_{2s}$ at the 
instant of time $t$.

 Let us  define the {\it current multiplicity} of the couple of vertices $\{\b,\g\}$, $\b,\g \in \bV_g$ up to the instant $t$ 
 by the following variable
$$
\fm_\CW^{(\{\b,\g\})}(t) = \# \{t' \in [1,t]: (\CW_{2s}(t'-1),\CW_{2s}(t'))= (\b,\g)\ \ 
{\hbox{or}} \ \  (\CW_{2s}(t'-1),\CW_{2s}(t'))= (\g,\b) \}
$$
and say that $\fm_\CW^{(\{\b,\g\})}(2s)$ represents the total multiplicity of the couple $\{\b,\g\}$.

The probability law of $ a_{ij}$ being symmetric, the weight of $\CI_{2s}$  (1.16)
is  non-zero
if and only if  
 $\CI_{2s}$ is such that in the corresponding graph of the walk  $\CW^{(\CI_{2s})}_{2s}$ each couple 
 $\{\a,\b\}$ has an even multiplicity
 $\fm_\CW^{(\{\a,\b\})}(2s) = 0( \hbox{mod}\, 2)$.
 We refer to the walks of such trajectories as to  the {\it even closed  walks} \cite{SS1} and denote
 by $\bW_{2s}$ the set of all possible even closed walks of $2s$ steps. In what follows, we consider
 the even closed walks only and refer to them simply as to the walks.


\vskip 0.2cm 
 It is natural to say that the pair $(\CW_{2s}(t-1), \CW_{2s}(t))= \fs_t$
represents the {\it step of the walk} number $t$.  
Given $\CW_{2s}\in \bW_{2s}$, we say that the instant of time $t$ is {\it marked} \cite{SS1} if the couple
$\{\a,\b\}= \{\CW_{2s}(t-1),\CW_{2s}(t)\}$
has an odd current multiplicity 
$\fm_\CW^{(\{\a,\b\})}(t)= 1(\hbox{mod}\, 2)$. We also say that  the corresponding step $\fs_t$   and the 
edge $e_t$
of $g(\CW_{2s})$
are   marked. All other steps and edges are called the {\it non-marked} ones.
Regarding  a collection of the marked edges $\bar \bE_g$ of $g(\CW_{2s})$, we consider a multigraph
$ \bar g_s = (\bar \bV_g, \bar \bE_g)$. Clearly, $\bar \bV_g = \bV_g$ and $\vert \bar \bE_g\vert=s$.
It is useful to keep the time labels of the edges $\bar \bE_g$ as they are  in  $\bE_g$.
Given two edges $e' = e_{t'}$ and $e''= e_{t''}$  such that $t'<t''$,  we write that $e'<e''$. Sometimes we denote
$t'= t(e')$.
\vskip 0.2cm
In general, $\bar g(\CW_{2s})$ is a multigraph with multiple edges. Replacing the multiple edge by a simple one,
we get a new graph that we refer to as the {\it skeleton} $S_{\bar g}$ of the graph $\bar g$.

\vskip 0.2cm
Any even closed  walk $\CW_{2s}\in \bW_{2s}$
generates a sequence $\theta_{2s}$ of
$s$ marked and $s$ non-marked instants. Corresponding sequence of $2s$ signs $+$ and $-$   is known to encode a
Dyck path of $2s$ steps. We  denote by   $\theta_{2s}= \theta(\CW_{2s})$ the
Dyck path of $\CW_{2s}$  and 
say that $\th(\CW_{2s})$ represents the {\it Dyck structure} of $\CW_{2s}$.
\vs 
Let us denote by
 $\Theta_{2s}$ the set of all Dyck paths of $2s$ steps.
It is  known that $\Theta_{2s}$ is  in one-by-one correspondence with the set of all half-plane rooted trees
$\CT_s\in \bT_s$ constructed with the help of
$s$ edges.
The correspondence between 
$\Theta_{2s}$ and $\bT_s$ can be  established with the help of  the chronological run $\fR$ over the edges of $\CT_s$.
It is known that the  cardinalities of $\bT_s$, $s=0, 1, 2,...$ are given by the Catalan numbers , $\vert \bT_s\vert = \rt_s$ (1.7).
We  refer to the elements of $\bT_s$ as to the {\it Catalan trees}.
We consider the edges of the tree $\CT_s$ as the oriented ones in the direction away from  the root of $\CT_s$.  

Given a Catalan tree $\CT_s\in \bT_s$, one can label its vertices with the help of letters of $\CA$ according to 
$\fR_{\CT}$. The root vertex gets the label $\a_1$ and each new
vertex that has no label is labelled by the next in turn letter. 
We denote the walk obtained  by
${\stackrel{\circ}{\CW}}_{2s} [\CT_s]$ and the corresponding Dyck path 
$\th_{2s} = \th({\stackrel{\circ}{\CW}}_{2s})$ will be denoted also as
 $\theta_{2s} = \theta(\CT_s)$.

Any Dyck path $\theta_{2s}$ generates   a sequence $(\xi_1, \xi_2, \dots, \xi_s)$, $\xi_i\in \{1,2, \dots, 2s-1\}$
such that each  step  $\fs_{ \xi_i}$, $1\le i\le s$  of ${\stackrel{\circ}{\CW}}_{2s}[\theta_{2s}]$ is marked.
We denote this sequence by $\Xi_s= \Xi(\th_{2s})$.  Given $\Xi_s$ and $\t\in [1, s]$,
one can uniquely reconstruct $\theta_{2s}$ and find corresponding instant of time $\xi_\t\in \{1,\dots, 2s-1\}$.
We will  say that the interval $[1,s]$ represents the {\it $\t$-marked instants} or
{\it instants of marked time} that varies from $1$ to $s$; sometimes we will  simply say that
$\t\in [1,s]$ is the {\it marked instant} when no confusion  can arise.

\vs

Given a walk $\CW_{2s}$ and a letter $\b$ such that $\b\in \bV_g(\CW_{2s})$,
we say that the instant of time $t'$ such that $\CW_{2s}(t')=\b$ 
represents an {\it arrival } $\fa$ at $\b$.
If $t'$ is marked, we will say that the corresponding arrival $\fa(\b)$ is the marked arrival at $\b$.
In $\CW_{2s}$, there can be several marked arrival instants of time at $\b$ that we denote by 
$1\le t^{(\b)}_1< \cdots < t^{(\b)}_N$.
For any non-root vertex $\b$, we have $N=N_\b\ge 1$. 
The first arrival instant of time $\b$ is always the marked one.
We can say that $\b$ is created at this instant of time.
To unify the description, we assume that the
root vertex $\varrho$
is  created at the  zero instant of time $t^{(\rho)}_1=0$ and add the corresponding zero  marked instant to the list of the
marked arrival instants at $\varrho$.

If $N_\b\ge 2$, then we say that the $N$-plet $(t^{(\b)}_1,\dots, t^{(\b)}_N)$ of marked arrival instants of time represents the
 {\it self-intersection} of $\CW_{2s}$, $\b$ is the {\it vertex of self-intersection}, and this self-intersection is of the {\it degree}
 $N$ \cite{SS1}.
 We say that the self-intersection degree  $\vk(\b)$ is equal to $N$ and denote this by  $\vk(\b)=N_\b$.
 If $N_\b = 1$, then we will say that $\vk(\b)=1$. 
 
 Finally, let us consider a vertex $\b$ and a collection of the marked edges  of the form $(\b,\a_i)$. We say
 that this collection is the {\it exit cluster} of $\b$ and denote it by $\D(\b)$,
 $$
 \D(\b) = \D_\CW(\b) = \{ e \in \bar \bE(\CW_{2s}):\ e=(\b,\a_i)\}.
 \eqno (2.5)
 $$
Sometimes we will say that $\D(\b)$ is given by the collection of corresponding vertices $\a_i$.

\vskip 0.2cm

Finally, let us determine the tree-type walks and (2,4)-walks we will study.
Given  $\CW_{2s}$, we can say that it is of the {\it tree-type form} if
$\bar g(\CW_{2s})$ has no cycles, i.e. when $\bar g(\CW_{2s})$ represents a tree when the multiple edges are considered  as the single ones.
If $\bar g(\CW_{2s})$ has at least one cycle, we can say that corresponding $\CW_{2s}$
is not of the tree-type structure. 
Regarding a walk of the tree-type form $\hat \CW_{2s}$, we will say that it is a (2,4)-walk if the weight
of the corresponding trajectory $\Pi(\CI_{2s})$ (2.2) contains the factors $V_2$ and $V_4$ only. 
If the (2,4)-walk is such that in its graph the multiple edges with the weights $V_4$ have no
vertices in common, we will say that this walk is the $(2,4)^*$-walk.

To complete this section, let us note that the number of the tree-type walks of $2s$ steps
whose weight is given by $V_{2}^{s}$ is given by the Catalan numbers $\rt_{s}$ (1.7). It follows 
from (1.1) that these numbers verify the following recurrent relation 
$$
\rt_k = \sum_{j=0}^{k-1} \rt_{k-1-j} \, \rt_j, \quad k\ge 1
\eqno (2.6)
$$
with the initial condition $\rt_0 = 1$. In the studies of the $(2,4)$-walks, we find a number of 
generalizations of relations (1.7) and (2.6).

\section{Walks of non-tree type}

\subsection{Classification of  vertices and  weights of walks}

Given a walk $\CW_{2s}$ and its graph $\bar g(\CW_{2s})=(\bV_g, \bar \bE_g)$ and consider  two vertices 
  $\a$ and $\b$. We denote by 
$ \CE_{\{\a,\b\}}$ the collection of all marked edges of the form $(\a,\b)$ or $(\b,\a)$ of $\bar g(\CW_{2s})$ and determine the minimal edge $\tilde e = \min\{e: \,  e \in \CE_{\{\a,\b\}}\}$.
Let us assume that $\tilde e$ is of the form $(\a,\b)$.

    If  
 the multi-edge $\CE_{\{\a,\b\}}$ contains the edge $e_1$ of the first arrival at $\b$,  
 $e_1 = e(\fa_1(\b))$,
 then $\tilde e = e_1$ and 
 we say  that this edge $\tilde e =(\a,\b)$ {creates} $\b$
and that $\tilde e$ is the {\it base edge of} $\b$ or simply the base edge.

Let us consider the edge of the second arrival at $\b$, $e_2 = e(\fa_2(\b))$.
If $e_2 = (\a,\b)$, then we color it in green and say that $\b$ is the {\it green $p$-vertex}. 

Let us consider the edge $e_2 =e(\fa_2)= (\g,\b)$ of the second arrival at $\b$ such that $\g\neq \b$. 
If $e_2$ is the minimal edge of the multi-edge  $\CE_{\{\b,\g\}}$, 
then we   say that $\beta$ is the {\it blue $r$-vertex} and color 
$e_2$ in blue.

Let us consider the case when 
$\tilde e = \min \{ e: \, e\in \CE_{\{\g,\b\}}\}= (\b,\g)$.  
 If $\tilde e$ is the edge of the first or the second 
arrival at $\g$, $\tilde e = \fa_i(\g)$ with $i=1$ or $i=2$, 
then we color $e_2= (\g,\b)$ in red and say that $\b$ is the {\it red $q$-vertex}. 
If $\tilde e = \fa_j(\g)$ with $j\ge 3$,
then we color  $e_2= (\g,\b)$ in blue 
and consider $\b$ as the blue $r$-vertex.

It is not hard to see that all edges of the second arrival to one or another vertex
are colored and that their colors are uniquely determined. 
All remaining edges of $\bar g(\CW_{2s})$ that are not the base or the color ones are referred as to the {\it grey $u$-edges}.

\vskip 0.2cm
{\bf Lemma 3.1.} {\it Let $\CI_{2s}$ be such that the graph of its walk $\CW_{2s}= \CW(\CI_{2s})$ contains $r$ blue $r$-vertices, 
$p$ green $p$-vertices, $q$ red $q$-vertices. Also we assume that $\bar g(\CW_{2s})$ has $u$ grey  $u$-edges. Then the weight of $\CI_{2s}$  (2.2) is bounded as follows
$$
\Pi_a(\CI_{2s}) \, \Pi _b(\CI_{2s})= \Pi_a(\CW_{2s}) \, \Pi _b(\CW_{2s}) \le \left( { V_2^2\over n^2}\right)^r \ \left( { V_2 U^2\over n\r}\right)^{p+q} 
\left( { U^2\over \r}\right)^{u} \ \left( {V_2\over n}\right)^{s-u -2(r+p+q)}.
\eqno (3.1)
$$
}

\vskip 0.1cm {\it Remark.} To make the statements of the present section and their proofs clearer, 
we keep the factors $V_2$ as they are remembering that $V_2=1$. 

\vskip 0.2cm
{\it Proof of Lemma 3.1.} 
The weight of the walk $\Pi_{a,b}(\CW_{2s}) = \Pi_a(\CW_{2s}) \, \Pi _b(\CW_{2s}) $ is given by the product
of weights of all existing multi-edges $\Pi_{a,b}(\CE_{\{\d,\epsilon\}})$. It is easy to see that the 
weight of the multi-edge can be estimated as follows,
$$
\Pi_{a,b}(\CE_{\{\d,\epsilon\}} )\le \left({ V_2\over n} \right)^{I_{b,r}} \left( { U^2\over \r}\right)^{I_{r,p} + I_q + u(\CE) },
$$
where $I_{b,r} = 1$ if the minimal edge $\tilde e $ of $\CE_{\{\d,\epsilon\}}$ is either the base one or the blue one and zero otherwise,
$I_{r,p} $ is equal to one if $\CE_{\{\d,\epsilon\}}$ contains a green edge and zero otherwise, 
$I_{q} $ is equal to one if $\CE_{\{\d,\epsilon\}}$ contains a red edge and zero otherwise
and $u(\CE) = u(\CE_{\{\d,\epsilon\}})$ represents the number of the grey edges in $\CE_{\{\d,\epsilon\}}$. 
Due to this factorization,  the weight of the walk can be estimated by the product of factors $V_2/n$ and $U^2/\r$ that
can be rearranged in the product with respect to  all vertices of the graph $g(\CW_{2s})$. This can be done by attributing
the weights $V_2/n$ to all base and blue edges  of the graph $\bar g(\CW_{2s})$ and the weights $U^2/\r$ to all green, red and grey edges
of  $\bar g(\CW_{2s})$ and by attributing to each vertex $\b$ the product of  weights of all edges that enter $\b$. 
It is clear that any color vertex has exactly one edge of the second arrival that is of the same color as the vertex. This observation completes the proof of Lemma 3.1. 
$\Box$

\vskip 0.2cm 

\subsection{Tree-type walks and walks of non-tree type}

Given a walk $\CW_{2s}$, we say that is it a {\it tree-type walk} if its graph $\bar g(\CW_{2s})$
does not contain any blue $r$-vertex. We denote by $\hat \bW_{2s}$ a collection of tree-type walks. 
If $\CW_{2s}$ is such that its graph $\bar g(\CW_{2s})$ contains at least one blue $r$-vertex,
then we say that $\CW_{2s}$ is of {\it non-tree type}. We denote a collection of all non-tree-type
walks by $\tilde \bW_{2s}$.

The following simple statement plays an important role in our studies.
\vskip 0.2cm 

{\bf Lemma 3.2.} {\it If $\CW_{2s}$ is such that  its graphical representation $g(\CW_{2s})$ 
has at least one red $q$-vertex, then $g(\CW_{2s})$ contains at least one blue $r$-vertex. }

\vskip 0.2cm We prove Lemma 3.2 in Section 5.  
 Regarding the example walk $\CW_{16}$ (2.3), we see that its graphical presentation contains 
two vertices of the self-intersection degree 2 (these are $\a_1$ and $\a_4$) and one vertex
$\a_2$ of the self-intersection degree 3.
Among vertices of $\bar g(\CW_{16})$, there is one $p$-vertex 
$\a_4$ and one $q$-vertex $\a_2$. The root vertex $\a_1$ has one blue edge of the (mute) first arrival
and one blue edge $e(6)$ of the second distinct arrival. So, the root vertex $\a_1$ is the blue $r$-vertex
and $\CW_{16}$ is of non-tree type.

\vskip 0.2cm
According to definitions, we can rewrite relation (2.1) in the form 
$$
M_{2s}^{(n,\r)} = \tilde \CZ_{2s}^{(n,\r)} + \hat \CZ_{2s}^{(n,\r)},
\eqno (3.2)
$$
where 
$$
\tilde \CZ_{2s}^{(n,\r)} = \sum_{\CI_{2s}: \  \CW^{(\CI_{2s})} \in \tilde \bW_{2s}}  \Pi_a(\CI_{2s}) \, \Pi_b(\CI_{2s})
$$
and 
$$
\hat \CZ_{2s}(n,\r) = \sum_{\CI_{2s}: \  \CW^{(\CI_{2s})} \in \hat  \bW_{2s}}  \Pi_a(\CI_{2s}) \, \Pi_b(\CI_{2s}).
$$
The following statements represents the main technical result of this section. 

\vskip 0.2cm

{\bf Theorem 3.1.} {\it 
Under conditions of Theorem 1.1, the following relation holds
$$
\tilde \CZ_{2s_n}(n,\r) = 
o(n\rt_{s_n}) , \quad (n,s,\r)\to\infty.
\eqno (3.3)
$$
} 

\vs {\it Remark.} Observing that all  terms of the right-hand side of the definition of $\hat \CZ_{2s}^{(n,\r)}$  are non-negative,
we conclude that 
$\hat \CZ_{2s}(n,\r) \ge n \rt_s V_2^s= n\rt_s$. Therefore relation (3.3) implies the asymptotic estimate
$$
\tilde \CZ_{2s}(n,\r) = o( \hat \CZ_{2s}(n,\r)), \quad (n,s,\r)\to\infty.
\eqno (3.4)
$$


\subsection{ Diagrams $\CG^{(c)}(\bar \nu)$ and their realizations }

Each walk $\CW_{2s}$ generates a set of numerical data, $\bar \nu = (\nu_2, \nu_3,\dots, \nu_s)$,
where $\nu_k$ is the number of vertices $\b_i$ of $\bar g(\CW_{2s})$ such that their self-intersection degree is equal to $k$, 
$\vk(\b_i) = k$. 
To estimate the number of elements of the set 
$\tilde \bW_{2s}$, we  construct a kind of  diagrams 
$\CG^{(c)} (\bar \nu)$.

\vskip 0.2cm

To explain general principles of the estimates, let us start with   the construction of {\it non-colored} 
 diagram $\CG(\bar \nu)$.
This diagram  consists of $\vert \bar \nu \vert = \sum_{k=2}^s \nu_k$ 
vertices. We arrange these vertices  in $s-1$ levels, the $k$-th level contains  $\nu_k$
vertices. Each vertex $v$ of $k$-th level is attributed by $k$ half-edges that have heads attached to $v$ but have 
no tails. Instead of the tail of each edge, we join  a square box (or window) to it.
Then  any vertex $v$ of this $k$-th level 
has $k$ edge-boxes (or edge-windows) attached. 

Given $\CG(\bar \nu)$, one can attribute to its edge-windows 
the values from the set $\{1,2, \dots, s\}$ such that there is no pair of windows with the same value. 
The diagram together with the corresponding values produces 
a realization of $\CG(\bar \nu)$ that we denote by 
$\la \CG(\bar \nu)\ra_s$.

\vskip 0.2cm

The principal observation used  by the Sinai-Soshnikov method  
is  that 
an even closed walk $\CW_{2s}$ can be completely  determined by its values 
at  the marked instant of time added by a family of rules that 
indicate  the values of the walk at  the non-marked instant of time. 
Given Dyck path  $\th_s$ and a realization $\la \CG(\bar \nu)\ra_s$,
the positions of the 
walk at the marked instants of time are  completely determined.

The values at the non-marked instant of time are determined
by a family of rules $\bY(\bar \nu)$ that 
indicate the way to leave a vertex $\b$ of self-intersection
with the help of the non-marked step out. 
It is shown in \cite{SS1,SS2} that if $\varkappa(\b)=k$,
then the number of the exit rules  at this vertex is bounded as follows,
$\vert \bY(\b)\vert \le (2k)^k$. The rigorous proof of this upper bound is given in \cite{KV} (see also \cite{K}). 
No such rule as $\U$ is needed for the
non-marked instants of time when the walk leaves a vertex of the self-intersection degree 1 because in this case 
the continuation 
of the run is uniquely determined. 
Then the total number of the rules can be estimated as follows,
$$
\vert \bY(\bar \nu )\vert \le \prod_{k=2}^s \, (2k)^{k\nu_k}.
\eqno (3.5)
$$

The number of all possible realizations of $\CG(\bar \nu)$
is given by the following expression
$$
\sum_{ \la \CG(\bar \nu)\ra_s} 1 = 
{s!\over \nu_2! (2!)^{\nu_2}\,  \nu_3! (3!)^{\nu_3}\cdots \nu_s! (s!)^{\nu_s} \cdot (s-\Vert \bar \nu \Vert)!},
$$
where $\Vert \bar \nu \Vert = \sum_{k=2}^s k\nu_k$. 
It is easy to see that 
 the following upper bound is true, 
$$
\sum_{ \la \CG(\bar \nu)\ra_s} 1 \le 
\prod_{k=2}^s {1\over \nu_k!} \left({s^k\over k!}\right)^{\nu_k}.
$$
Combining this inequality with  (3.5), we conclude that the number of elements in
$\bW_{2s}(\bar \nu)$ can be estimated as follows,
$$
\vert \bW_{2s}\vert \le \rt_s \prod_{k=2}^s {1\over \nu_k!} \left({(2k)^k \, s^k\over k!}\right)^{\nu_k}
\le  \rt_s \prod_{k=2}^s {(C_1s)^{k\nu_k}\over \nu_k!}\,  ,
\quad {\hbox{where }}\quad C_1 = \sup_{k\ge 1} {(2k+2)\over (k!)^{1/(k+1)}}.
\eqno (3.6)
$$
We have introduced the constant $C_1$ in the form that simplifies further computations.

\vskip 0.2cm

The upper bound  (3.6) clearly explains the role of the diagrams 
$\CG(\bar \nu)$ in the estimates of the number of walks. However, it is rather rough and does not give 
 inequalities needed in the majority of cases of interest. In particular, the estimate (3.6) is hardly  
 compatible with the upper bound of the weight 
 of walks (3.1) in the case of dilute random matrices. 

\vskip 0.2cm 
To improve the upper bound (3.6), we adapt the diagram technique to our model by introducing 
 more informative   diagrams based on $\CG(\bar \nu)$. Also, we formulate a new filtering principle
to estimate more accurately the number of walks. 
A kind of the filtration  principle has been implicitly  used already by Ya. Sinai and A. Soshnikov. 
The rigorous formulation of the filtration technique is given in \cite{K}. In paper \cite{K3} it 
was adapted to the study of the moments of dilute random matrices.

\vskip 0.5cm
Let us describe the construction of the {\it color diagram} $\CG^{(c)}(\bar \nu, \bar p, \bar q)$  determined by parameters
$\bar \nu = (\nu_2, \dots, \nu_s)$, $\bar p= (p_2, \dots, p_s)$
and $\bar q= (q_2, \dots, q_s)$. We start with the non-colored diagram $\CG(\bar \nu)$ and 
consider 
 $\nu_k$ vertices of the $k$-th level of it.  
We fill  the second edge-box attached to each vertex by using the   set 
$\{1,\dots,s\}$.
This can be done by 
$$
{s!\over \nu_k! (s-\nu_k)!} \le {s^{\nu_k} \over  \nu_k!} = {s^{r_k+p_k+q_k} \over  \nu_k!}
\eqno (3.7)
$$
ways. Then we color the $\nu_k$ vertices in blue, red and green colors by one of 
$
{\nu_k!\over r_k!  \, p_k!\, q_k!}
$
ways, where $r_k = \nu_k - p_k-q_k$. 
Then we color corresponding edge-boxes in grey, blue, red and green colors. 
The base edge-boxes of the first arrivals  attached to blue vertices are colored in blue.
Instead of boxes, the base edges of the red and green vertices get circles colored 
with respect to the color of the corresponding vertex.

Taking  the empty $k-2$ edge-boxes attached to  green or red vertex,
we fill them with the values from the set $\{1,\dots, s\}$. 
This can be done by not more than $s^{k-2}/(k-2)!$ ways. 
Regarding the edges of the first arrivals at red and green vertices that remain empty,
we replace corresponding boxes by circles colored according to the color of the vertex.

Let us consider  $k-1$ empty edge-boxes attached to a blue vertex
and fill them with the values from $\{1,\dots, s\}$. 
Ignoring the  restriction of the edge-box of the second arrival, we estimate the number of 
ways to do this  by expression $s^{k-1}/(k-1)!$.

This procedure being performed at each level independently,
we get the following estimate from above of the number of different realizations of 
color diagrams,
 $$
\sum_{\la \CG^{(c)}(\bar \nu, \bar  p, \bar q)\ra_s} 1\  \le \ 
\prod_{k=2}^s \ {1\over r_k!} \left( {s^k\over (k-1)!} \right)^{r_k} 
\cdot {1\over p_k!} \left( { s^{k-1}\over (k-2)!}\right)^{p_k}
\cdot 
{1\over q_k!} \left( { s^{k-1}\over (k-2)!}\right)^{q_k}.
\eqno (3.8)
$$

\vskip 0.2cm 
The {\it filtration procedure} is follows: we consider a realization of the color diagram $\la\CG^{(c)}\ra_s$
such that  all grey, blue, red and green boxes of edge-windows of $\CG^{(c)}$ are filled with different values of 
$\{1,\dots, s\}$ while the red and green circles of the first arrivals at the $q$-vertices and $p$-vertices remain empty.

Having a Dyck path $\theta_s$ and a rule $\U\in \bY(\bar \nu)$ pointed out, 
we   start the run of the  walk $\CW$ according $\theta_s$, $\la\CG^{(c)}\ra_s$ and $\U$
  till the marked instant of the first $p$-edge  or $q$-edge appear. Let us denote by $v\rq{}$ the corresponding
vertex of the diagram $\CG^{(c)}$. 
Let us denote 
the marked instant mentioned above by $\t'$ with $t'= \xi_{\t'}$
  and assume that the sub-walk $\CW_{[0,t'-1]}$ get its end value $\b = \CW_{[0,t'-1]}(t'-1)$. 
  Then at the instant of time $t'$ the walk has to choose one of the admissible vertices 
from the set $\Gamma=\{\g_1, \dots, \g_L\}$
such that the edge $(\b, \g_j)$ possesses the properties of either $p$-edge or $q$-edge, respectively.
Clearly, the set $\Gamma$ depends on the color of the edge-box with $\t\rq{}$. 
Once the vertex $\gamma_j$ is chosen, we take the marked instant of the first arrival at $\gamma_j$
and record its value to the edge-box of the first arrival $\CO_1(v\rq{})$. Clearly, the number of walk
is bounded by $\vert \Gamma\vert$. This is why it is natural 
to say that we apply the filtering of all possible values to fill $\CO_1(v\rq{})$. 

Having chosen the value of $\CO_1$, we continue the run of the walk, if it is possible,
till the marked value of the second arrival at the next in turn red or green vertex $v\rq{}\rq{}$ is seen.
Then the filtering procedure is repeated. 
 When all the walk is constructed, if it exists, we denote 
by $\la \la \CG^{(c)}\ra^{(b)}_s\ra_{\CW}$ the set of values
in red and green circles obtained during this run of $\CW$. 

\vskip 0.2cm
{\bf Lemma 3.3.} {\it Given a realization of a color diagram   $\la \CG^{(c)}(\bar \nu, \bar p, \bar q)\ra_s$,
let us denote by 
$\bW_{2s} (D,  \la \CG^{(c)}(\bar \nu, \bar p, \bar q)\ra_s, \U)$ 
the set of walks $\CW_{2s}$ that have this  realization of $\CG^{(c)}$ and follow the rule $\U$  
and such that the maximal exit degree
$$
\CD(\CW_{2s}) = \max_{\b \in \bV(\CW_{2s})}\vert \D(\b)\vert
$$ 
is equal  to $D$, $\CD(\CW_{2s})=D$.
 Then 
the number of possible realizations of the values of red and green circles of $\CG^{(c)}$
admits the following upper bound, 
$$
\vert \la \la \CG^{(c)}\ra_s\ra_{\CW}\vert \le  
\, 2^{\vert \bar q\vert } \, D^{\vert \bar p\vert}, 
$$
 where $\vert \bar q\vert = \sum_{k=2}^s q_k$ and $\vert \bar p\vert = \sum_{k=2}^s p_k$
 and therefore
 $$
 \vert \bW_{2s}(D, \la \CG^{(c)}(\bar \nu, \bar p, \bar q)\ra_s, \U) \vert 
 \le \, 2^{\vert \bar q\vert } \, D^{\vert \bar p\vert}\, \rt_s .
\eqno (3.9)
 $$
 
 }

{\it Proof of Lemma 3.3}. 
Let us prove (3.9) in the case when the color diagram $\CG^{(c)}$ has the only one red vertex $v$. 
Following the filtration principle, we take a Dyck path $\th_s$ and perform the run of the walk in accordance with 
 the 
data  given by the self-intersections of $\la \CG^{(c)}(\bar \nu, \bar p, \bar q)\ra_s^{(b)}$ and $\U$
till the value $\xi_{\t'}$ appear, where $\t'$ is attributed to the second arrival edge-box attached to  $v$.
By  the definition, the edge $(\CW(\xi_{\t\rq{}}-1),\CW(\xi_{\t\rq{}}-1))= (\b, \g)=e$
is red only in the case when the edge $\tilde e = (\g,\b)$ is the edge either of the first or the 
second arrival at $\g$ and $\tilde e < e$. Therefore the sub-walk $\CW_{[0, \xi_{\t\rq{}}-1]}$
has not more that two vertices available to join at the instant $\xi_{\t\rq{}}$.  This explains the
factor 2 in the left-hand side of (3.9).

In the case when $v$ is the only one green vertex, the sub-walk $\CW_{[0, \xi_{\t\rq{}}-1]}$
has the set $\D(\b)$ completely determined, and the vertex to join at the instant
$\xi_{\t\rq{}}$ necessarily belongs to this set.
The upper bound (3.9) in the general case can be proved by recurrence.
 $\Box$

\vskip 0.3cm
Now we can  estimate  the number of walks
that have a color diagram $\CG^{(c)}(\bar \nu, \bar p, \bar q)$ and the maximal exit degree $D$, 
 $$
\vert \bW_{2s}(D,  \CG^{(c)}(\bar \nu, \bar  p, \bar q)) \vert \le \rt_s
\prod_{k=2}^s \  { ( C_1 s)^{kr_k}\over r_k!} 
\cdot { \left( D (C_1s)^{k-1}\right)^{p_k}\over p_k!}
\cdot 
{ \left( { 2(C_1s)^{k-1}}\right)^{q_k} \over q_k!}.
\eqno (3.10)
$$
This relation  follows from  
inequalities (3.5), (3.8) and  (3.9).

 \vs
 We will use Lemma 3.3 and a version of relation (3.10) in the proof of Theorem 2.1 below. 
 However, to get the estimates we need, we have to show that the number of Catalan trees $\CT_s$
  generated by the elements of $\bW_{2s}(D,\CG^{(c)})$ is exponentially small with respect to the 
 total number $\rt_s$ of all $\CT_s$  \cite{SS2,S}.  To do this, we need to study the vertex of maximal exit degree
 of walks $\CW_{2s}$ in more details.

 \subsection{Vertex of maximal exit degree, arrival cells and BTS-instants}

 Let us consider a walk $\CW_{2s}$ and find the first letter 
that we denote by  $\bb$  such that
$$
 \vert \D(\bb)\vert = \CD(\CW_{2s}).
 \eqno (3.11)
$$
We will refer to $\bb$  as to the {\it vertex of maximal exit degree} and
 denote for simplicity $D= \CD(\CW_{2s})$.
 To classify the arrival instants at $\bb$, we 
need to determine
 reduction procedures  similar to those considered in  \cite{KV} and further modified in \cite{K3}.
 Certain elements of the reduction procedure of \cite{KV} were independently introduced in paper \cite{FS}.


\subsubsection{Reduction procedures and reduced sub-walks}

 Given $\CW_{2s}$, 
let  $t' $ be the minimal instant of time such that 
\vs 
\hskip 0.4cm i) the step $\fs_{t'}$ is the marked step of $\CW_{2s}$;

\hskip 0.4cm ii) the consecutive to $\fs_{t'} $ step $\fs_{t'+1}$ is non-marked;

\hskip 0.4cm iii) $\CW_{2s} (t'-1) = \CW_{2s}(t'+1)$.
\vs
\noindent 
If such $t'$ exists, we  apply to the ensemble of steps $\fS = \{ \fs_t, 1\le t\le 2s, \fs_t \in \CW_{2s}\} $
a reduction  $\dot \CR$ that removes from $\fS$ two consecutive elements $\fs_{t'}$ and $\fs_{t'+1}$; 
we denote $\dot \CR(\fS) = \fS'$. The ordering time labels of elements of $\fS'$ are inherited from 
those of $\fS$.

\vs 

The new sequence $\fS'$  can be regarded again as an even closed walk. We can apply to this new walk the reduction procedure
 $\dot \CR$. 
Repeating  this operation   maximally possible number of times $m$,
 we get the walk
$$
\dot \CW_{2\dot s} = ( \dot \CR)^{m}
(\CW_{2s}), \quad \dot s = s-m,
$$
that we refer to as  the {\it strongly reduced walk.} We denote $\dot \fS = (\dot \CR)^m(\fS)$ and say that $\dot \CR$
is the {\it strong reduction} procedure.

\vs 

We introduce  a {\it weak reduction} procedure $\breve R$ of $\fS$ that removes
from   $\fS_{2s}$
the pair  $(\fs_{t'}, \fs_{t'+1})$ in the case when  the conditions (i)-(iii) are verified
and

iv) $\CW_{2s}(t')\neq \bb$.

\noindent We denote by 
$$
\brW_{2\breve s}= (\breve \CR)^{l}(\CW_{2s}), \quad  \breve s = s-l
\eqno (3.12)
$$
 the result of the action
of maximally possible number of consecutive weak reductions $\breve \CR$ and denote $\breve \fS = (\breve \CR)^l (\fS)$. 
In what follows, we sometimes omit the subscripts $2\dot s$ and $2\breve s$.
 Regarding the example walk $\CW_{16}$ (2.3), we observe that $\bb = \a_3$ and 
  that the strongly and weakly reduced walks coincide
and are as follows,
$$
\dot \CW_8 = \breve \CW_8 = (\a_1, \a_2, \a_3, \a_5, \a_1, \a_2, \a_3, \a_5, \a_1).
$$

 \vs
Taking the difference $\breve \fS\setminus \dot \fS = \ddot \fS$, 
 we see that it represents  a collection
of sub-walks, $\ddot W = \cup_j \ddot \CW^{(j)}$.
 Each sub-walk $\ddot \CW^{(j)}$  can be reduced by a sequence of  the strong reduction procedures $\dot \CR$
 to an empty walk. We say that $\ddot \CW^{(j)}$ is of the {\it Dyck-type} structure.
  It is easy to see that 
any $\ddot \CW^{(j)}$ starts by a marked step and ends by a non-marked steps and there is
no steps of $\dot W$ between these two steps of $\ddot \CW^{(j)}$.  
We say that $\ddot \CW^{(j)}$ is the {\it non-split} sub-walk.

\vs
It is not hard to see that the collection of steps $\check \fS = \fS\setminus \breve \fS$ is given by a collection of 
subsets $\check \fS = \cup_k \check \fS^{(k)}$, each of $\check \fS^{(k)}$ represents a  non-split Dyck-type sub-walk
$\check \CW^{(k)}$, 
$$
\check W = \cup_k \check \CW^{(k)}.
\eqno (3.13)
$$
In this definition,  we assume that each sub-walk   $\check \CW^{(k)}$ is maximal by its length.

\subsubsection{Arrival instants and Dyck-type sub-walks attached to  $\bb$}

Given $\CW_{2s}$, let us consider the  instants of time $0\le t_1< t_2<\dots t_R\le 2s$
such that for all $i=1,\dots, R$ the walk  arrives at $\bb$ by the steps
of  $\brW_{2\breve s }$,
 $$ 
\CW_{2s}( t_i) = \bb \quad {\hbox{and}} \quad
\fs_{t_i}\in \brW_{2\breve s}, \quad i=1,2,\dots, R.
\eqno (3.14)
$$
We say that  $t_i$ are  the {\it $\breve t$-arrival} instants of time of  $\CW_{2s}$.
Let us consider   a sub-walk 
that corresponds to the subset $\fS_{[t_i+1,t_{i+1}]} = \{\fs_t, t_i+1\le t\le t_{i+1}\}  \subseteq\fS$; we denote this sub-walk by 
$\CW_{[t_i, t_{i+1}]}$.  In general, we denote a sub-walk that is not necessary even and/or closed  by $\CW_{[t',t'']}$ also.

\vs Let us consider  the interval of time $[t_i+1, t_{i+1}-1]$ between two consecutive $\breve t$-arrivals at $\bb$.
It can happen that $\CW_{2s}$ arrives at $\bb$ at some instants of time $t' \in [t_i+1,t_{i+1}-1]$, $\CW_{2s}(t') = \bb$.   
We denote by  $\check  t_{(i)}$ the maximal value of such $t'$. 

\vs
{\bf Lemma 3.4.} {\it The sub-walk $\CW_{[t_i, \check  t_{(i)}]}$  coincides 
with one of the maximal Dyck-type
sub-walks $\check  \CW^{(k')}$ of (3.13). }

Lemma 3.4 is proved in \cite{K3}.

\vs

 Let us consider
a collection  of all marked exit edges from $\bb$ performed  by the marked steps on the interval
of time 
$[t_i, \tilde t_{(i)}]$ and denote this collection by $\check \D_i$.
We
say that $\check  \D_j$ represents the {\it exit sub-clusters of Dyck type} attached to  $\bb$.
or simply that $\check \D_j$ are the {\it exit sub-clusters} of $\CW_{2s}$. We denote
their cardinalities by 
 $d_j = \vert \check  \D_j\vert$.
 The exit sub-clusters are ordered  in natural way.
 To keep a unified description, we accept the existence of empty exit sub-clusters; then 
 we get equality $D =\sum_{j=1}^R d_j, d_j\ge 0$. 
 Clearly, any exit sub-cluster is attributed to  a uniquely determined $\bt$-arrival instant at $\bb$.

Regarding the reduced walk $\brW_{2\breve s}$ of $\CW_{2s}$ (3.12), we can determine corresponding 
Dyck path $\breve \theta_{2\breve s} = \theta (\brW_{2\breve s})$ and the tree
 $\breve \CT_{\breve s}
= \CT(\breve  \theta)$.
It is easy to show  that $\breve  \CT_{\breve s}$
is  a sub-tree of  the original tree $\CT_s = \CT( \theta(\CW_{2s}))$. One can introduce  the difference
$\check \CT = \CT_s \diagdown
 \breve \CT_{\breve s}$ and say that  it  is represented by a collection of
sub-trees $\check  \CT^{(j)}$. 

\vs 
Returning to the Catalan tree $\CT(\th_{2s})$,   let us consider  
the chronological run over it that we denote by $\fR_\CT$. Then  the 
$\breve t$-arrival instant  $t_l$ (3.14) determines 
the  step $\varpi_l$ of $\fR_\CT$. 
Also the corresponding vertex $\check \u_l$ of the tree $\CT_s$ is determined.
It is clear  that $\check \u_l$ are not necessarily different for different $l$. 

The sub-trees $\check \CT^{(l)}$ are attached to  $\check \u_l$
and the chronological run over $\check \CT^{(l)}$ starts immediately after
the step $ \varpi_l$ is performed.
We will say that these steps  $ \varpi_l$, $1\le l\le R$ represent   
the {\it nest cells} from where  the sub-trees
$\check  \CT^{(l)}$, $1\le l\le L$ grow. It is clear that the sub-tree $\CT_l$ has
$d_l\ge 0$ edges attached to its root $\vr_l$ and this root coincides with  the vertex $\check \u_l$.
Returning to $\CW_{2s}$, we will say that the arrival instants of time $\breve t_l$
represent the {\it arrival cells} at $\bb$.
In the next sub-section, we describe a classification of the arrival cells at $\bb$ 
that represents a natural improvement of the approach proposed in \cite{KV}.

\subsubsection{Classification of arrival cells at $\bb$}

Let us consider a walk $\CW_{2s}$ together with its reduces counterparts $\dot \CW_{2\dot s} = \dot \CW$ 
and $\breve \CW_{2\breve s} = \breve \CW$. Let  $t_i$ denote a $\bt$-arrival cell (3.4). 
If the step  $\fs_{t_i}$ of $\CW_{2s}$ is marked,
then we say that $t_i$ represents  a {\it  proper cell} at $\bb$.
If the step  $\fs_{t_i}$ is non-marked and  $\fs_{t_i}\in \ddot  \CW= \breve \CW\setminus \dot \CW$,
then we say that $t_i$ represents a {\it mirror cell} at $\bb$.
If  the step $\fs_{t_i}\in \dot \CW$ is non-marked,
then we say that $t_i$ represents an {\it imported cell } at $\bb$.

Let  us consider $I$ proper cells  $\ddot t_i$ 
such that  $\fs_{\ddot t_i}$ belongs to $\ddot \fS$. We denote by 
$ x_i$ the corresponding marked instants, $ x_i = \xi_{\ddot t_i}$,
$1\le i\le  I$ and write that $\bar x_I = (x_1, \dots, x_I)$.
  It is easy to see that each proper cell $x_i$ can be attributed by   a number  1 or 0 in dependence of whether
 it produces a corresponding mirror cell at $\bb$ or not. 
 We denote this number by $m_i \in \{  0,1\}$ and write that $M= \sum_{i=1}^{I} m_i$ and 
 $\bar m_I = (m_1, \dots, m_I)$. Clearly, $M\le I$.

Regarding the strongly reduced walk $\dot \CW_{2\dot s}$, we denote by 
 $\dot t_k$ the proper cells such that the steps $\fs_{\dot t_k} \in \dot \fS$. Corresponding
 to $\dot t_k$ marked instants will be denoted by $z_k$, $1\le k\le  K$. Then 
 $\bar z_K = (z_1,\dots, z_K)$ and 
 clearly $\vk(\bb) = I+K$.

Given  $\CW_{2s}$ with non-empty  
 set  $\dot \fS$,  there exists at least one pair of elements of $\dot \fS$ denoted by 
$(\fs', \fs'')$ such that $\fs'$ is a marked step of $\dot \CW_{2\dot s}$, $\fs''$ is the non-marked one 
and $\fs''$ follows immediately after $\fs'$ in $\dot \fS$. We refer to each pair of this kind
as to  the pair of {\it broken tree structure steps} of $\CW_{2s}$ or in abbreviated form, 
 the {\it BTS-pair} of $\CW_{2s}$. 
If $\tau' $ is the marked instant that corresponds to $\fs'$, 
we will simply say that $\tau'$ is the {\it BTS-instant}  of $\CW_{2s}$ \cite{KV}.

Regarding the strongly reduced walk $\dot \CW$, let us  consider 
   a non-marked arrival step at $\bb$ that we denote by $\bar \fs = \fs_ {\bar t}$.
   Then one can find a uniquely determined   marked instant 
    $\t'$ such that all steps  $ \fs_t\in \dot \fS$
with $\xi_{\t'}+1\le t \le \bar t$ are the non-marked ones.
Let us denote by $ t''$ the instant of time of the first non-marked step $\fs_{\bar t''} \in \hat \fS$
of this series of non-marked steps. 
Then $(\fs_{t'}, \fs_{ t''})$ with  $t' = {\xi_{\t'}}$  is the BTS-pair of $\CW_{2s}$ that corresponds to $\bar t$.
We will say that $\bar t$ is attributed to  the corresponding BTS-instant $\t'$. 
It can happen that several arrival instants $\bar t_i$ are attributed to  the same BTS-instant $\t'$.
We will also say  that the BTS-instant $\t'$ {\it generates}  the imported cells that are attributed to it.

\vs  

Let us consider a BTS-instant $\t$   such that 
$\CW_{2s}(\xi_{\t}) =  \bb$.
As it is said above, we  denote such marked instants by $z_k$, $1\le k\le K$. 
Assuming that a marked BTS-instant $z_k$ generates $f_k'\ge 0$ imported cells, we denote
by $\vp^{(k)}_1, \dots, \vp^{(k)}_{f'_k}$ the positive numbers   such that 
$$
\CW_{2s}( \xi_{z_k} + \sum_{j=1}^l \vp^{(k)}_j) = \bb
\quad {\hbox{for all}} \quad 1\le l\le f'_k.
\eqno (3.15)
$$ 
If for some $\tilde k$ we have $f'_{\tilde k}=0$,
then we will say that $z_{\tilde k}$ does not generate any imported cell at $\bb$.
We denote $\bar \vp^{(k)} = (\vp^{(k)}_1, \dots, \vp^{(k)}_{f'_k})$. 
\vs 

Let us consider a BTS-instant $\t$ that generates imported cells at $\bb$ and such that 
$\CW_{2s} (\xi_\t)\neq \bb$. We denote these BTS-instants by $y_j$, $1\le j\le J$. 
Assuming that a marked BTS-instant $y_j$ generates $f''_j+1$ imported cells, $f_j''\ge 0$,
we denote
by $\L_j, \psi^{(j)}_1, \dots, \psi^{(j)}_{f''_j}$ the positive numbers   such that 
$\CW_{2s} (\xi_{y_j} + \L_j) = \bb$ and 
$$
\CW_{2s}\left( \xi_{y_j} +\L_j+  \sum_{l=1}^k \psi^{(j)}_l\right) = \bb \quad {\hbox{for all}} \quad 1\le k\le f''_j.
\eqno (3.16)
$$ 
In this case we will say that the first arrival at $\bb$ given by the instant of time
$\xi_{y_j} + \L_j$ represents the {\it principal} imported cell at $\bb$.
All subsequent arrivals at $\bb$ given by (3.13) represent the {\it secondary} imported cells at $\bb$. 
We will say that $y_j$ is the {\it remote} BTS-instant with respect to $\bb$ and will use denotations  
$\bar y_J= (y_1, \dots, y_J)$ and $\bar \L_J = (\L_1, \dots, \L_J)$.
We also denote 
$\bar \psi^{(j)}= (\psi^{(j)}_1, \dots, \psi^{(j)}_{f''_j})$.

We see that for a given walk $\CW_{2s}$, the proper, mirror and imported cells 
at its vertex of maximal exit degree
are characterized by  the  set of parameters, $(\bar x, \bar m)_{ I}$, 
$(\bar z, \Phi,\bar f')_{K}$, where 
$\Phi_K= (\bar \vp^{(1)}, \dots \bar \vp^{(K)})$, $\bar f'_{K}  = (f'_1, \dots, f_{K}')$ and 
$(\bar y, \bar \L,  \Psi, \bar f'')_J$, where 
$\Psi_J= (\bar \psi^{(1)},\dots, \bar \psi^{(J)})$, 
$\bar f''_J = (f''_1, \dots, f''_J)$. We also denote
$$
F' = \sum_{k=1}^{K} f'_k \quad {\hbox{ and}}\quad   
F'' = \sum_{j=1}^{J} f''_j.
$$
Summing up, we observe that 
the  vertex  $\bb$ with  $\varkappa(\bb) = I+K$
has the  total number of cells  given by $R = I+M+K+ J+F$,  where 
$F=F'+F''$. In what follows, we  denote the family of the parameters described above by
$$
\CP_R= \left\{ (\bar x,\bar m)_I, (\bar y, \bar \L,  \Psi, \bar f'')_J, (\bar z, \Phi,\bar f')_{K}\right\}.
\eqno (3.17)
$$

\subsection{Proof of Theorem 3.1}

We are going to estimate the number of walks in a family  of walks 
$\tilde \bW_{2s}(D)$ that have a vertex of maximal exit degree $D$.
We rewrite (3.2) in the following from
$$
\tilde \CZ_{2s}(n,\r) = \sum_{D=1}^s \, \sum_{\CW_{2s} \in \tilde \bW_{2s}(D)} \, \Pi_a(\CW_{2s} ) \,  \Pi_b(\CW_{2s}) \cdot
\vert \CC_{\CW_{2s}}\vert ,
$$
where $\vert \CC_{\CW_{2s}}\vert $ is given by (2.4). To estimate the number of elements in $\tilde \bW_{2s}(D)$, we have to 
consider a  kind of color diagrams that have a separate vertex $\breve v$ attributed by the parameters
from the family $\CP_R$, namely by $\bar x_I$ and $\bar z_K$. Also we have to include into the color diagrams 
the parameters $\bar y_J$. Let us describe this new type of color diagrams.

\subsubsection{Color diagrams with a vertex of maximal exit degree}

Let us consider a vertex $\breve v$  and  attach to it  $I+K$ edge-boxes. 
We  denote by $\la \breve v_{I,K}\ra_s$ a realization
of the values of marked instants that fill   these boxes.
Given $\bar \nu, \bar p$ and $\bar q$, we  consider a realization  of the corresponding color diagram 
$\la \CG^{(c)}(\bar \nu, \bar  p, \bar q)\ra_s$ and point out 
$J$ edge-boxes that will provide the marked instants 
 $\bar y$. 
Joining  such a realization with  chosen $J$ edge-boxes   
$\la \CG^{(c)}_{\bar y}(\bar \nu, \bar  p, \bar q)\ra^{(b)}_s$ with $\la \breve v_{I,K}\ra_s$,  
we get a realizations of the  diagram 
we need, 
$$
 \la\breve  \CG^{(c)}_{\bar x, \bar z, \bar y}(\bar \nu, \bar  p, \bar q)\ra^{(b)}_s = \la \breve v_{I,K}\ra_s 
\uplus \la \CG^{(c)}_{J}(\bar \nu, \bar  p, \bar q)\ra^{(b)}_s = \la \breve v \uplus \CG^{(c)}\ra_s^{(b)}.
$$
The last equality of the formula presented above introduces  a denotation
for a realization of the diagram we consider. 

The number of different realizations of the color diagram $\CG^{(c)}(\bar \nu, \bar p, \bar q)$ is estimated
by the right-hand side of (3.8).  Regarding  realizations $\la \breve v_{I,K}\ra_s$,
we can write that 
$$
\vert \la \breve v_{I,K}\ra_s\vert \le {s^{I+K}\over (I+K)!} 2^{I+K},
\eqno (3.18)
$$
where the last factor gives the upper bound for the choice of $K$ elements among $I+K$ ones
to be marked as the values of  $\bar z_K$. The vertex  $\breve \b$ of the walk can be attributed 
by the weight
$$
\Pi_a(\bb) \, \Pi_b(\bb) =  \begin{cases}
{ \displaystyle  V_2\over  \displaystyle n}, & \text{if $\vk(\bb)=1$}, \\
{\displaystyle 1\over \displaystyle n^2\r^{I+K-2}}V_2^2 U^{2(I+K)-4} , & \text{if $\breve \b$ is an $r$-vertex}, \\
{\displaystyle 1\over  \displaystyle  n\r^{I+K-1}} V_2 U^{2(I+K)-2} , & \text {if $\bb$ is a $p$-vertex or a $q$-vertex.}
\end{cases}
\eqno (3.19)
$$
In the first and in the third  cases of (3.19), 
at least one  blue $r$-vertex is necessarily present in $\CG^{(c)}(\bar \nu,\bar p,\bar q)$.

\vs
Regarding $\la\CG^{(c)}(\bar \nu,\bar p,\bar q)\ra_s$, one can choose $J$ edge-boxes
to be labeled as the values of the realization $\la \bar y\ra$ among 
$\sum_{k=2}^s (k-1)\nu_k = \Vert \bar \nu \Vert_1$ edges only.
This is because the first arrival to a vertex cannot be the marked BTS-instant. 
 The number of ways to choose $J$ ordered places among $\Vert \bar \nu\Vert_1$ ordered edges
can be estimated as follows,
$$
{\Vert \bar \nu\Vert_1 \choose  J} \le { (\Vert \bar \nu\Vert_1)^J\over J!} \le 
{1\over h_0^J} \, \exp\left\{ h_0 \Vert \bar \nu\Vert_1\right\}, 
\eqno (3.20)
$$
where $h_0>1$ is a constant.

\subsubsection{Exit sub-clusters and cells at $\bb$}

The maximal exit degree of a walk  $\CW_{2s}\in \bW_{2s}(D) $
can be represented as follows,  $D = \dot  D +  \ddot D+  \check D$, where $\dot D$ is the number of marked edges
of the form $(\bb, \g)$ that belong to the strongly reduced walk $\dot \CW$ (3.12), $\ddot D$ represents the exit edges
that belong to $\ddot \CW = \breve \CW \setminus \dot \CW$.
It is known that $\dot D = F+J$  and that $F\le K$ \cite{KV} (see also Lemma 5.1 of \cite{K}). 
Also we observe that $\ddot D = M$.  
Taking into account that $M\le I$,  we can write that 
$$
\check D = D - M-F-J \ge D - I-K-J.
\eqno (3.21)
$$

\vs
The remaining  $\check  D$ edges of $\bar \bE(\CW_{2s})$ 
belong to the exit sub-clusters of the Dyck-type sub-walks $\check \CW^{(k)}$ (3.13)
attached to $\bb$. They are distributed among $R$ arrival cells at $\bb$. 
We denote by $\bar d = (\check d_1, \dots, \check d_R)$ a particular distribution such that
$\sum_{l=1}^R \check d_l =\check D$.

The number of cells $R$ depends on  $\la \CG^{(c)}\ra_s^{(b)}$, $\th_s$ and $\U$. 
However, the inequalities used to get (3.21) show that 
$$
R = I+K+M+F+J\le 2I +2K+J  = R^*.
\eqno (3.22)
$$ 
Then 
$$
\sum_{\bar d_R, \vert \bar d_R\vert = \check D} 1 = { \check D + R-1\choose R-1} \le {D+R^*-1\choose R^*-1}.
$$
Elementary analysis shows that if $D\ge 2$, then 
$$
{D+R^*-1\choose R^*-1}  \le 
h_0^{R^*} \sup_{R^*\ge 2} {1\over h_0^{R^*-1}} {D+R^*-1\choose R^*-1} \le 
h_0^{2I+2K+J} \exp\left\{ { eD\over h_0} \right\}, \ h_0>e.
\eqno (3.23)
$$
Indeed, using the well-known estimates 
$$
\sqrt{2\pi n} \left( {n\over e}\right)^n\le n!\le e \sqrt n \left( { n\over e}\right)^n, \quad n\ge 1,
$$
we can write that 
$$
{1\over h_0^m} {(D+m)!\over D!\, m!}\le 
{1\over h_0^m} {e\over 2\pi} \sqrt {{D+m\over mD}}\, 
\left( 1 +{m\over D}\right)^D\, \left(1+ {D\over m}\right)^m\le 
{e^m\over h_0^m} \left( 1+ {D\over m}\right)^m,
$$
where we take into account that $D+m\le 2mD$. Then the last relation of (3.23) follows. 
Now we are ready to perform the estimates that prove    Theorem 3.1.

\subsubsection{Exponential estimates and $\tilde \CZ_{2s}$}

In this subsection we estimate the contribution of the non-tree type walks $\tilde \CZ_{2s}$ and 
 prove relation (3.3) with the help of computations that are very similar to those 
used in the pioneering papers by Ya. Sinai and A. Soshnikov.  The following statement can be regarded as 
the principal result in the rigorous formulation of the method.

\vs
{\bf Lemma 3.4.}  {\it Given $D$, a realization of the color diagram $\la \breve v\uplus \CG^{(c)}\ra^{(b)}_s$ (3.18)
and a rule $\U$,
let us consider a family of walks $\bW_{2s} (D, \la \breve v\uplus \CG^{(c)}\ra_s, \U)$ such that 
the vertex of the maximal exit degree $\bb$ given by  $\breve v$  has $D$ exit edges of the form
$(\bb,\g_i)$.
Then
$$
\vert \bW_{2s} (D, \la \breve v\uplus \CG^{(c)}\ra_s, \U)\vert \le 
2^{\vert \bar q\vert } \, D^{\vert \bar p\vert } \,
 \left( e^{\eta} h_0^2\right)^{I+K+ J}\,  e^{ - \eta D + eD/h_0} \, \rt_s,
\eqno (3.24)
$$ 
where $\eta = \ln (4/3)$.}

\vs We prove Lemma 3.4 in Section 5. 
The walks we consider are of the non-tree type and therefore contain at least one blue
$r$-vertex $v_0$. Let us divide the sum $\tilde \CZ_{2s}$ in two parts in dependence whether 
$v_0=\breve v$ or $v_0\neq \breve v$,
$$
\tilde \CZ_{2s}(n,\r) = \tilde \CZ_{2s}^{(1)}  + \tilde \CZ_{2s}^{(2)},
\eqno (3.25)
$$
respectively.
Then we can write that  
$$
\tilde \CZ_{2s}^{(1)} 
= \sum_{D=1}^s \ \sum_{J=0}^s \,
\left( \prod_{k=2}^s \, \sum_{r_k, p_k, q_k}\right)^\star 
\ \sum_{I,K: \, I+K\ge 1}\ 
 \vert \CC_{W_{2s}}\vert
$$
$$\times \sum_{  \breve v\uplus \CG^{(c)}_s} \ 
\sum_{ \la \breve v\uplus \CG^{(c)}\ra_s} \ 
\sum_{ \CW_{2s} \in \bW_{2s} (D, \la \breve v\uplus \CG^{(c)}\ra_s)}\ 
\Pi_{a,b}(\CW_{2s}) ,
\eqno (3.26)
$$
where the star means that the values of $r_k, p_k$ and $q_k$ are such that $\sum (k-1)\nu_k\ge J$ and  $\sum r_k\ge 1$. 
The first sum of the second line of (3.26) takes into account the choice of the $J$ places in $\CG^{(c)}$ to be marked
as the edge-boxes of values $y_j$ (see also (3.20)); the second sum is performed over all possible realizations of the 
diagram $\breve v\uplus \CG^{(c)}$ with the help of the values from $\{1,\dots, s\}$ (see (3.8) and (3.18)).
\vskip 0.2cm

Using relations (3.1), (3.10), (3.18) and (3.19), we deduce from (3.26) the following upper bound, 
$$
\tilde \CZ^{(1)}_{2s} \le 
\sum_{D=1}^s \ \sum_{J=0}^s \ \sum_{I+K\ge 1} {(2s)^{I+K}\over(I+K)!} \left(2(I+K)\right)^{I+K} \cdot 
h_0^{2(I+K)+J}\,  e^{\eta (I+J+K) }\ 
$$
$$
\times  \left( \prod_{k=2}^s\,  {\sum_{r_k, p_k, q_k}}\right)^\star
  e^{ h_0 \Vert \bar \nu\Vert_1} 
\  {  1\over r_k!} \left( { (2k)^k s^k\over (k-1)!}\right)^{r_k}
\cdot {1\over p_k!}  \left( {(2k)^k D s^{k-1}\over (k-2)!}\right)^{p_k} 
\cdot {1\over q_k!} \left( {(2k)^k 2s^{k-1}\over (k-2)!}\right)^{q_k}
$$
$$
\times  e^{-\eta D + eD/h_0}\, \rt_s \cdot n^{s+1  - (\Vert \bar \nu\Vert_1 +(I+K-1))}
$$
$$
\times\
{V_2 U^{2(I+K-1)}\over n \, \r^{I+K-1}}  \cdot \left( {V_2^2\over n^2}\right)^{r_k}  \, 
\left( { V_2 U^2\over n\r}\right)^{p_k+q_k} \, \left( {U^2\over \r}\right)^{(k-2)\nu_k}
\, \left( {V_2 \over n}\right)^{s-\Vert \bar \nu \Vert-(I+K)} ,
\eqno (3.27)
$$
where we denoted $\Vert \bar \nu \Vert = \sum_{k=2}^s \, k\nu_k$.
\vs 
Let us consider a constant (cf. (3.6))
$$
C_2= \max\left\{ \sup_{k\ge 2} {2k\over ((k-1)!)^{1/k}},\, 
\sup_{k\ge 2} {(2k)^{k/(k-1)}\over ((k-2)!)^{1/(k-1)}}, 
\right\}
$$
and denote 
$$
B= C_2 h_0 e^{h_0 +\eta} = 4C_2h_0e^{h_0}/3,
$$
where $h_0>e$ will be determined below.
Remembering that $s= \chi \rho$, we can deduce from (3.27) the following inequality,
$$
\tilde Z^{(1)}_{2s} \le V_2^s  \sum_{D=1}^s    e^{-\eta D + eD/h_0} n\, \rt_s \ \sum_{J=0}^s \sum_{I+K\ge 1} 
2s B \left( 2B \hat U^2\chi\right)^{I+K-1} 
$$
$$
\times  \left( \prod_{k=2}^s\,  {\sum_{r_k, p_k, q_k}}\right)^\star
 \  {  1\over r_k!} \left( {B\hat U^2 s^2\over n} (B\hat U^2\chi)^{k-2}\right)^{r_k}
\cdot {1\over p_k!}  \left( D (B\hat U^2\chi)^{k-1}\right)^{p_k} 
\cdot {1\over q_k!} \left( 2(B\hat U^2\chi)^{k-1}\right)^{q_k}.
\eqno (3.28)
$$
If $\chi $ is such that 
$$
2B\hat U^2 \chi \le 1,
\eqno (3.29)
$$
then (3.29) implies inequality
$$
\tilde Z^{(1)}_{2s}\le 4Bs^3 \left( \exp\left\{ {2Bs^2\over n}\right\} -1\right)
\cdot e^{4B\hat U^2\chi} \, n \rt_s V_2^s  \ 
\sum_{D=1}^\infty \exp\left\{-\eta +2B\hat U^2\chi + {e/ h_0} D\right\}.
\eqno (3.30)
$$
Remembering that $\eta = \ln (4/3) > 0.28$, we see that if  
$$
{3C_2 U^2 h_0 e^{h_0}\over V_2}\chi + {e\over h_0}\le 0.28
\eqno (3.31)
$$  
then  (3.30) 
implies relation
$$
\tilde Z^{(1)}_{2s} = o(n \rt_s V_2^s), \quad n, \r_n\to\infty
$$
provided $\r_n = o(n^{1/5})$ (1.6). Clearly, the  choice of $h_0$ and $\chi$ such that
$$
h_0 = 4e \quad {\hbox {and }} \quad \chi \le {V_2\over 400  e^{4e+1} C_2 U^2}
\eqno (3.32)
$$
makes (3.29) and (3.31) valid. Let us note that 
more detailed analysis of the walks with maximal exit degree $D$  show that the factor $s^3$ in the right-hand side of (3.30) can be 
eliminated. However, in the present paper we do not aim  the maximal rate of $\r_n$ and therefore the upper bound (3.30) is
sufficient for our purposes.

\vskip 0.2cm
Let us consider the second term of (3.25). The sub-sum $\tilde \CZ_{2s}^{(2)}$ can be estimated from above
by the expression given by the right-hand side of (3.27), where the sum over $I,K$
is performed over the range $I+K\ge 2$ and the weight factor  $V_2 U^{2(I+K-1)}/(n\r^{I+K-1})$
is replaced by $V_2^2 U^{2(I+K-2)}/(n^2 \r^{I+K-2})$ (see relation (3.20))
and where the condition $\sum_k r_k\ge 1$ is omitted.

Then we can write that
$$
\tilde Z^{(2)}_{2s} \le n \rt_s V_2^s  
\sum_{D=1}^s    e^{-\eta D + eD/h_0} \ \sum_{J=0}^s \sum_{I+K\ge 2} 
{2s^2 B\over n} (2B\hat u^2\chi)^{I+K-2}
$$
$$
\times   \prod_{k=2}^s\,  {\sum_{r_k, p_k, q_k}}
 \  {  1\over r_k!} \left( {B\hat U^2 s^2\over n} (B\hat U^2\chi)^{k-2}\right)^{r_k}
\cdot {1\over p_k!}  \left( D (B\hat U^2\chi)^{k-1}\right)^{p_k} 
\cdot {1\over q_k!} \left( 2(B\hat U^2\chi)^{k-1}\right)^{q_k}.
\eqno (3.33)
$$
If (3.32) is true, then  (3.33) implies the upper bound
$$
\tilde Z^{(2)}_{2s} \le n \rt_s V_2^s {4 s^4 B\over n} \cdot e^{2B\hat U^2  \chi} 
\ \sum_{D=1}^\infty 
\exp\left\{-\eta +2B\hat U^2\chi + {e/ h_0} D\right\}.
$$
Then
$$
\tilde Z^{(2)}_{2s} = o( n \rt_s V_2^s)
$$
under conditions of Theorem 2.1. Theorem 3.1 is proved.
 $\Box$



\section{Tree-type walks and $(2,4^\star)$-walks}

Let us consider the family $\hat \bW_{2s}$  of tree-type walks and  separate it into two  non-intersecting subsets,
$$
\hat \bW_{2s} = \dot \bW_{2s} \sqcup \ddot \bW_{2s},
$$
where $ \dot \bW_{2s}$  contains the walks $\CW_{2s}$ such that their weights have the factors $V_2=1$ and $V_4$ only and the graph 
$\bar g(W_{2s})$ is such that the $V_4$-edges do not share  a vertex in common.
We also denote this set by $\bW^{(2,4^\star)}$ and say that if $\CW_{2s}\in \bW^{(2,4^\star)}$,
then this $\CW_{2s}$ is the tree-type {\it $(2,4^\star)$-walk}.
 We denote
$$
\dot  \CZ_{2s}(n,\rho) = \sum_{ \CW_{2s}\in \dot \bW_{2s} }  \Pi_{a,b}(\CW_{2s}) 
\cdot \vert \CC_{\CW_{2s}}\vert  
\quad {\hbox{and}}\quad  
\ddot \CZ_{2s}(n,\rho) = \sum_{\CW_{2s}\in \ddot \bW_{2s} }  
\Pi_{a,b}(\CW_{2s}) \cdot \vert \CC_{\CW_{2s}}\vert. 
$$
Let us point out that two following relations are true, 
$$
 \vert \CC_{\CW_{2s}}\vert = n^{\vert \bV(\CW_{2s})\vert}(1+ o(1)), \ n\to\infty
\quad {\hbox{and }}\quad \vert \CC_{\CW_{2s}}\vert \le n^{\vert \bV(\CW_{2s})\vert},
\eqno (4.1)
$$
where $\bV(\CW_{2s})$ is the ensemble of vertices of the graph $\bar g(\CW_{2s})$.

\vskip 0.5cm 

{\bf Theorem 4.1.} 

{\it Under conditions of Theorem 2.1, the following upper bounds are true}

$$
\limsup_{(n,s,\r)\to\infty} {1\over nt_s } \dot  \CZ_{2s}^{(n,\rho)}\le 4 \exp\{ 16 V_4  \chi  \}
\eqno (4.2)
$$
{\it and }
$$
\limsup_{(n,s,\r)\to\infty} {\rho \over nt_s} \ddot  \CZ_{2s}^{(n,\rho)}\le {C\chi } \exp\{ 16 V_4 \chi \},
\eqno (4.3)
$$
f{\it or all $0<\chi\le \chi_0= \chi_0(U)$ and $C\ge C_0 = C_0( U)$, where 
$$
\chi_0( U) = {1\over 4^{11} U^2}
\quad {\hbox{and}}\quad 
C_0(U)  = 3\cdot 4^{16} U^6.
\eqno (4.4)
$$   
}

{\it Remark. } Theorem (4.1) can be proved under conditions of Theorem 1.1 with (1.5) replaced by 
much less restrictive condition on the probability distribution of $a_{ij}$ to be such that all its moments
exist and are bounded as follows,
$$
V_{2+2k} \le k!\,V_2\,   b_0^k , \quad  k=2,3,\dots
\eqno (4.5)
$$
with given $b_0>0$ (see also \cite{K01}). In this case the constants of (4.4) should be  replaced by
$$
\chi_0\rq{}(b_0)=  {1\over 3\cdot 2^{19} b_0}
\quad {\hbox{and }} \quad 
C^\prime_0( b_0) =  3\cdot 4^{16} b_0^2,
\eqno (4.6)
$$
respectively, where we assumed that (4.5) holds with $V_2=1$.

\vs 

To describe the general structure of the tree-type walk, let us introduce several auxiliary notions. 
Regarding a sub-walk of $2a$ steps $\CW_{2a}$ and its graph $\bar g(\CW_{2a})$, let us 
denote by $\G_\vr$ the ensemble of the multiple edges of $\bar g$ that make a connected component
attached to the root $\vr$. If the graph $\bar g(\CW_{2a})$ has no other multiple edges
than those of $\G_\vr$ and the first step and the last step of $\CW_{2a}$ are performed along the 
edges of $\G_\vr$, we say that   $\CW_{2a}$ is the element of  the block of the first level 
$\bB_{a}^{(1)} (\G)$, $\G= \G_\vr$,
$$
\CW_{2a} = \CB_{a}\in \bB_a^{(1)}(\G).
$$
We will say also that $\CW_{2s}$ by itself is a block of the first level, when no confusion can arise.

We say that a walk $\CW_{2b}$ is a block of the second level, $\CW_{2s}= \CB^{(2)}_{b}$,
if it starts and ends with the steps along the root component of multiple edges $\G_\vr$
and there in $\CW_{2b}$ there exists at least on sub-walk $\CW_{2a}\rq{}$ that is the block of the first level. 

By recurrence, we say that $\CW_{2s}$ belongs to the block of the $(k+1)$-th level,
if it starts and end by the steps of $\G_\vr$ and contains sub-walks $\CB^{(l_1)}, \dots, \CB^{(l_p)}$
such that $l_i\le k$ and there exists at least one $l_j$ such that $l_j=k$. 

In general, the tree-type walk $\CW_{2s}$ is such that its graph $\bar g (\CW_{2s})$ represents 
a tree $\CT_h$ of $h\ge 0$ edges and along the chronological run over $\CT_h$, the 
sub-walks $\CB^{(l_1)}_{a_1}, \dots, \CB^{(l_q)}_{a_p}$ of the levels $1\le l_j\le s/3$ appear at the different moments $t_1, \dots t_q$, 
$t_j\in [0, 2h+1]$, $t_i\neq t_j$.

\vskip 1cm
\subsection{Proof of Theorem 4.1} 

Let us introduce the weight
$$
\pi(\CW_{2s}) = \prod_{e\in S_g} {V_{2m_e}\over\r^{m_e-1}},
$$
where $S_{\bar g}$ is the skeleton of the graph $\bar g(\CW_{2s})$ and $m_e$ is the multiplicity of the edge $e$. 

\vskip 0.2cm

{\bf Lemma 4.1.}  {\it Let us consider a family of walks $\bW^\diamond(m)$ such that all edges of 
$\bar g(\CW_{2m})$ are multiple and all of them  are attached to the root $\vr$. Then for any given
$0<\chi \le \chi_0$ and $C\ge C_0$ (4.4) the following estimate,
$$
P(m) = \sum_{\CW_{2m}\in \bW^\diamond(m)} \pi(\CW_{2m}) \le 
{C\over 4^{4m} \, \r^{1 +{\rI}_{\{m\ge 3\}}}} 
\eqno (4.7)
$$
holds for all $1\le m \le s= \lfloor \chi \r\rfloor-1$, 
where $\rI_{A}=1$ if $A$ is true  and $\rI_A=0$ otherwise.}

\vskip 0.5cm
{\it Proof.} In the cases of  $m=2,3,4$ relation (4.7) can be deduced directly from relations 
  $$
P(2) = {V_4\over \r}, \quad P(3) = {V_6\over \r^2},  \quad
{\hbox{and}}\quad  P(4) = {3V_4^2\over \r^2}.
$$
In the general case of $m\ge 5$, we can write that 
$$
P(m)=  \sum_{l=1}^{m-5} {m-1\choose l} {V_{2+2l}\over \r^l} P(m-1-l)
+{m-1\choose 3}
 {V_{2m-6} V_6\over\r^{m-2}} 
+{m-1\choose 2} 
 {V_{2m-4} V_4\over \r^{m-2}} + {V_{2m}\over\r ^{m-1}}.
\eqno (4.8)
$$
Using  upper bound (1.5) and assuming that (4.7)
 holds for $P(m-1-l) $ of the right-hand side of (4.8),
we can write that 
$$
P(m) \le {C\over 4^{4m} \r^2} \cdot R,
$$
where 
$$
R= \sum_{l=1}^{m-5} {4^{4+4l }\over  l!}\cdot {V_2 U^{2l} s^l\over \r^l}
+{(m-1)(m-2)(m-3)\over 3!}\cdot {4^{4m} V_2^2 U^{2m-4}\over C \r^{m-4}}.
$$
$$
+  {(m-1)(m-2)\over 2} \cdot {4^{4m} V_2^2 U^{2m-4}\over C \r^{m-4}} 
+ {4^{4m} V_2 U^{2m-2}\over C \r^{m-3}}.
\eqno (4.9)
$$
Denoting $\phi = 4^4 U^2 \chi$ and remembering that $V_2=1$, we deduce from (4.9) that 
$$
R\le 4^4  \, \phi\,  e^\phi + {4^{16}  U^4\over C } \phi
\cdot  \max_{m\ge 5}  {m(m-1)(m-2)\over 6 m^{m-4}}  
+\phi^2 {4^{12}  U^4\over C m^{m-3}}.
$$
It is easy to see that under conditions of Lemma 4.1, $R\le R_0<1$.  
Similar computations based on (4.8) 
show that Lemma 4.1 is true under conditions (4.5) and (4.6). $\Box$

\vskip 0.5cm 

Given a Catalan  tree $\CT_h$, let us consider the ensemble of vertices $\u_1, \u_2,\dots, \u_q$
that have exit edges and denote by $\d_1, \d_2 \dots, \d_q$ the number of such edges, $\d_i\ge 1$. In this case we will 
say that 
 the tree $\CT_h$ has $q$ inner vertices
with exit clusters (or bushes) 
$\D_1, \dots, \D_q$ (cf. (2.5)).  Given $\bar \mu = (\mu_1, \dots, \mu_q)$
with $\mu_i\ge \d_i$,  let us consider the family of walks 
$\bW^\diamond(\CT_h,\bar \mu)$ of $2h+2\hat m$ steps, $\hat m = \mu_1+\cdots +\mu_q$ such that 
all edges of their graphs $\bar g(\CW_{2h+ 2\hat m})$ are multiple, $S_{\bar g}= \CT_h$
and the vertex $\u_i$ has $\d_i+\mu_i$ exit edges, $1\le i\le q$. 
\vskip 0.5cm

{\bf Lemma 4.2.}
{\it Given $\bar \mu = (\mu_1, \dots, \mu_q)$
with $\mu_i\ge \d_i$,   consider the family of walks 
$\bW^\diamond(\CT_h,\bar \mu_q)$ of $2h+2\hat m$ steps, 
$\hat m = \vert \bar \mu_q\vert =\mu_1+\cdots +\mu_q$ such that 
all edges of their graphs $\bar g(\CW_{2h+ 2\hat m})$ are multiple, the skeleton of $\bar g$ 
is given by  $S_{\bar g}= \CT_h$
and the vertex $\u_i$ has $\d_i+\mu_i$ exit edges, $1\le i\le q$. Then
under conditions of Lemma 4.1,
$$
P(\CT_h,\bar \mu_q)=  \sum_{\CW_{2h+2\hat m} \in  \bW^\diamond(\CT_h, \bar \mu_q)} \pi (\CW_{2h+2\hat m})
 \le 4^{h+ \hat m} \prod_{j=1}^q P(\d_j+\mu_j).
\eqno (4.10)
$$
 }

\vskip 0.2cm
{\it Proof.} We prove Lemma 4.2 by recurrence. On the first step,
let us consider the family of walks $\bW^\diamond(p,\mu_0)$ such that each walk
 $\CW_{2p+ 2\mu_0}$  has  the graph $\bar g(\CW_{2p+2\mu_0})$ with all edges attached to the root 
$\rho$, the skeleton $S_{\bar g}$ is a bush with $p$ edges and each  edge of $S_{\bar g}$ is multiple.
Then 
$$
\sum_{\CW_{2p+2\mu_0} \in \bW^\diamond(p,\mu_0) } \pi (\CW_{2p+2\mu_0}) \le P(p+ \mu_0)
\eqno (4.11)
$$
and (4.10) is true. Relation (4.11) follows from the obvious observation that 
$\bW^\diamond(p,\mu_0) \subset \bW^\diamond(p+\mu_0)$. 

On the next step of the recurrence, 
let us  consider the family of walks $ \bW^\diamond(p, \mu_0; \bar \d_p, \bar \mu_p)$ such that 
their skeleton is given by a tree that has a  bush with the root $\vr$ and $p$ exit edges $(\vr,\u_i)$,
$1\le i\le p$
and there are $\d_i>0$ main exit edges and $\mu_i$ additional exit edges at  each $\u_i$, $1\le i\le p$. 
Given a sub-walk $\CW_{2p+2\mu_0} \in \bW^\diamond(p,\mu_0)$ of (4.11), 
we denote by $\k_1, \dots , \k_p$ the multiplicities of the $p$ edges of its skeleton. 
It is not hard to see that 
$$
\sum_{\CW \in \bW^\diamond(p,\mu_0; \bar \d_p, \bar \mu_p) } \pi (\CW) =
\sum_{\CW\in  \bW^\diamond(p,\mu_0) }  \pi(\CW) \ 
\prod_{i=1}^p \left( 
\Xi^{(\k_i)}(\d_i+\mu_i) \sum_{\CW\in \bW^\diamond(\d_i, \mu_i)} \pi(\CW) 
\right),
\eqno (4.12)
$$
where $\Xi^{(\k_i)}(\d_i+\mu_i)$ denotes the number of possibilities to perform $\d_i+\mu_i$ exits
from the vertex $\u_i$ after $\k_i$ arrivals to $\u_i$. In (4.12), we do not indicate the number of steps
of corresponding walks that is obvious. 
Taking into account the upper estimate
$$
\Xi^{(\k_i)}(\d_i+\mu_i) = {\d_i+\mu_i+\k_i-1 \choose \k_i-1} \le 2^{\d_i+\mu_i+\k_i},
\eqno(4.13)
$$
and relation (4.11), we deduce from (4.12) that 
$$
\sum_{\CW \in \bW^\diamond(p,\mu_0; \bar \d_p, \bar \mu_p) } \pi (\CW) 
\le
\sum_{\CW\in  \bW^\diamond(p,\mu_0) }  \pi(\CW) \ 
\prod_{i=1}^p \left( 2^{\d_i+\mu_i+\k_i}
 \sum_{\CW\in \bW^\diamond(\d_i, \mu_i)} \pi(\CW) 
\right)
$$
$$
\le 2^{p+\mu_0}  P(p+\mu_0) \prod_{i=1}^p 2^{\d_i+ \mu_i} P(\d_i+\mu_i). 
\eqno (4.14)
$$
Remembering that  $h=p+\d_1+\cdot +\d_p$ and $\hat m = \mu_0+\mu_1+\cdots + \mu_p$, we see that (4.10) follows from (4.14). 
\vskip 0.2cm
Now it is clear how to proceed in the general case of the walks 
$\CW\in \bW^\diamond(\CT;\bar \mu)$ whose skeleton is given by a  tree $\CT_h$. 
It is sufficient to consider the chronological run over $\CT_h$ and to find the first inner vertex $\u_0$
such that there are no inner vertices among its children $\u_1, \dots, \u_p$. In other words, 
 the vertex $\u_0$ is such that all of the exit edges  $(\u_0, \u_i) $ are leaves and $u_i$ are the outer vertices.

Therefore  we can apply (4.14) to the corresponding sub-walks
of $\CW\rq{} \in \bW^\diamond(p,\mu_0; \bar \d_p,\bar \mu_p)$.
Then we can consider the vertex $\u_0$
as the outer one with respect to the reduced walk $\CW\setminus \CW\rq{}$ and repeat the reduction procedure
by recurrence.

We see that in this process the vertex $\u_0$ and the edges   $(\u\rq{},\u_0)$
are considered twice in the estimates of the form (4.13):  first in the role of $\k$ enters at $\u_i$ and then in the role of 
$\d+\mu$ exits from $\u\rq{}$. Therefore the base of the exponent 2 is replaced by 4 
in the final estimate (4.10). We do not present the detailed computations
because they repeat those of (4.12) and (4.14). $\Box$ 

\vskip 0.2cm

{\it Corollary of Lemma 4.2.} Using (4.7), we deduce from (4.10) that 
$$
P(\CT_h, \bar \mu_q) \le {C^q\over 4^{3(h+\hat m)} \, 
\r^{q+ \sum_{i=1}^q \rI_{\{\d_i+\mu_i\ge 3\}}}}
\eqno (4.15)
$$
for any $c\ge C_0(V_2,U)$.

\vskip 0.5cm

{\bf Lemma 4.3. } {\it Let us denote by $B_s^{(1)}$ the sum of weights of all walks of $2s$ steps that represent
the blocks of the first level. Then 
$$
B_s^{(1)} =\sum_{\CW_{2s}\in \bB_s^{(1)}} \pi(\CW_{2s})
= \dot B_s^{(1)} + \ddot B_s^{(1)} , 
$$
where 
$\dot B_s^{(1)}$ is the sum over all walks that have only one multiple edge, this edge is the $V_4$-edge 
 attached to the root,
$$
 B_s^{(1)} = {V_4 \over \r}  T^{(3)}_{s-2} ,
\eqno (4.16)
$$
where 
$$
\quad  T^{(3)}_{s-2}= \sum_{
\stackrel{a_1,a_2, a_3 \ge 0}{a_1+a_2+a_3=s-2}}\   \rt_{a_1}\, \rt_{a_2} \, \rt_{a_3} = \rt_s\,  {3s\over 2(2s-1)}.
\eqno (4.17)
$$
If $\r\ge C$, then  
$$
\ddot B_s^{(1)}\le  {C\over 120 \r^2}\, \rt_s.
\eqno (4.18)
$$
}

\vskip 0.2cm
{\it Proof.} Taking into account  the definition of the blocks of the first level 
and remembering that $V_2=1$, we can write that 
$$
B_s^{(1)} = \sum_{h=1}^{\lfloor s/2\rfloor } \sum_{\CT_h} 
\sum_{\hat m\ge 2h}\ 
\sum_{\stackrel{\mu_1,\dots, \mu_q\ge 1}{\mu_1+\cdots+\mu_q = \hat m}}\
P(\CT_h, \bar \mu_q) \, T_{s-h-\hat m}^{(2(h+\hat m)-1)}\, ,
\eqno (4.19)
$$ 
where the sum is taken over all possible values of $\hat m$ and $\mu_i\ge 1$ and
similarly to (4.17),
$$
T_{s-h-\hat m}^{(2(h+\hat m)-1)}
=  \sum_{
\stackrel{a_i \ge 0}{a_1+\cdots +a_{2(h+\hat m)-1}=s-h-\hat m}}\ \prod_{i=1}^{2(h+\hat m)-1} t_{a_i}
\le 4^{2(h+\hat m)} \rt_s. 
\eqno (4.20)
$$
The last inequality of (4.20) is proved in Section 5. Using (4.15), we deduce from (4.19) that 
$$
B_s^{(1)} \le 
\rt_s \ \sum_{h=1}^{s/2} \sum_{\CT_h} \ {C^q\over 4^{2h} \, \r^q}
\prod_{i=1}^q \ \left( 
\sum_{\mu_i=1}^\infty {1\over 4^{2\mu_i} \, \r^{\rI_{\{ \d_i+\mu_i\ge 3\}}}}\right) 
$$
$$
\le \rt_s \ \sum_{h=1}^{s/2} \sum_{\CT_h} \ {C^q \,  \over 4^{2h} \, \r^q}
\prod_{i=1}^q \ \left( 
{1\over 4^2 \r^{\rI_{\{ \d_i\ge 2\}} }} + \sum_{\mu_i=2}^\infty {1\over 4^{2\mu_i} \, \r^2}\right)
\le 
\rt_s \ \sum_{h=1}^{s/2} \sum_{\CT_h} \ {C^q  \over 4^{2h} \, (8\r)^q} \,
\prod_{i=1}^q \ 
{1\over  \r^{\rI_{\{ \d_i\ge 2\}} }}.
\eqno (4.21)
$$
Regarding the last expression, we observe that if $h=1$, then $q=1$ and $\d_1=1$ and we get 
the term (4.17) in this case, and therefore  $\dot B^{(1)}_s\le 3V_4 \rt_s/(4\r)$. 
For the remaining terms of the sum over $h\ge 2$, we observe that 
if $q=1$, then the tree $\CT_h$ is determined uniquely with  $\d_1\ge 2$. 
Therefore we can write that 
$$
\sum_{h\ge 2} \sum_{\CT_h} \ {C^q\over 4^{2h} \, (8\r)^q}
\prod_{i=1}^q \ 
{1\over  \r^{\rI_{\{ \d_i\ge 2\}} }}\le 
\sum_{h\ge 2} {1\over 4^{2h}} \left( {C\over 8\r^2} + (\rt_h-1) {C\over 8\r^2}\right) \le 
{C\over 120 \r^2}, 
$$ 
where we have used inequality $C/\r\le 1$. Relation (4.18) is proved. 
We prove relation (4.17)  in Section 5. Lemma 4.3 is proved. $\Box$

\vskip 0.2cm
{\it Corollary  of Lemma 4.3.} 
It follows from (4.16), (4.17) and (4.18) that if $\r\ge C$, then 
$$
B_s^{(1)}\le  {V_4\over  \r}\, \rt_s .
\eqno (4.22)
$$
\vskip 0.5cm
{\bf Lemma 4.4.} {\it Let us denote by 
$B^{(k)}_s$ the sum of the weights of walks that represent the blocks of the $k$-th level.
If $\r\ge C$,
then  the following upper bound holds,
$$
B^{(k)}_s \le   {\alpha\rt_s \over \r}\,    \hat \chi^{k-1},
\eqno (4.23)
$$
where $\a = V_4$ and $\hat \chi = 128 V_4\chi$ with $\chi$ and $C$ determined by (4.4). 
}

\vskip 0.2cm
{\it Proof.} We prove Lemma 4.4 by recurrence. The case of $k=1$ is verified directly with the help of  (4.22).
In the general case, we can write that 
$$
B^{(k+1)}_s \le  \sum_{p\ge 1} \ \sum_{a+ b_1+\cdots + b_p=s} 
B_a^{(1)} { 2a-1\choose p}  \ p B^{(k)}_{b_1}
\  \prod_{i=2}^p \left( B^{(1)}_{b_i}+ B^{(2)}_{b_i} +\dots + B^{(k)}_{b_i}\right),
\eqno (4.24)
$$ 
where $B^{(1)}_a $ counts the sub-walks  attached to the root and ${2a-1\choose p}$ gives the upper bound
of the possibilities to choose the instants to start the remaining sub-walks of $\CB^{(l)}_{b_i}$; among them
there is at least one block of the $k$-th level, we denote the corresponding sum by $B_{b_1}^{(k)}$. 

Using (4.22) and assuming that (4.23) can be applied to the right-hand side of (4.24), we obtain that 
$$
B^{(k+1)}_s \le \sum_{p\ge 1}\ \sum_{a+ b_1+\cdots + b_p=s}  \rt_a \rt_{b_1} \cdots \rt_{b_p}\ 
{\a \over \r} \cdot {(2s\a)^p\over (p-1)! \r^p} \ \hat \chi^{k-1} 
\left( \sum_{j=1}^k \hat \chi^{j-1}\right)^{p-1}
$$
$$
\le 
{\a\over \r} \rt_s {2\cdot 4^2 s\a\over \r}\sum_{p\ge 1} 
{1\over p!} \left( {8\a\chi \over  1-\hat \chi}\right)^{p-1} 
\le {\a\rt_s \over \r}  {\hat \chi^k} {\exp\{ 8\a\chi/(1-\hat \chi)\}\over 4} .
\eqno (4.25)
$$
It is easy to deduce from  (4.4) that $\hat \chi <1/2$ and $\exp\{16\a\chi\}<4$. 
Then $B^{(k+1)} \le \a \hat \chi^{k} \rt_s/\r$ and Lemma 4.4 is proved. $\Box$

\vskip 0.5cm
 
{\bf Lemma 4.5.} 
{\it Denote by $\ddot B^{(k)}_s$ the sum of the weights of walks
that represent the blocks of $k$-th level and such that some of its sub-walks contains 
two or more $V_4$-edges that share a vertex or at least one $V_{2l}$-edge with $l\ge 3$.
Then
$$
\ddot B^{(k)}_s \le {\b \rt_s \over \r^2}\, \hat \chi^{k-1}, \quad k\ge 1,
\eqno (4.26) 
$$
where $\b = C/120$.}
\vskip 0.2cm
{\it Proof.}
Relation (4.26) with $k=1$ is verified by  (4.18).
In the case of $k\ge 2$, we can use the denotations of (4.24) and  write that 
$$
\ddot B^{(k+1)} \le  \sum_{p\ge 1} \ \sum_{a+ b_1+\cdots + b_p=s} 
{ 2a\choose p} p \left(\ddot B_a^{(1)} B^{(k)}_{b_1} + B_a^{(1)} \ddot B^{(k)}_{b_1}\right) 
\prod_{i=2}^p \sum_{j_i=1}^k \
 \ B_{b_i}^{(j_i)} 
$$
$$
+ \sum_{p\ge 1} \ \sum_{a+ b_1+\cdots + b_p=s} 
{ 2a\choose p} p (p-1) B_a^{(1)} B^{(k)}_{b_1}  \left(\ddot B^{(1)}_{b_2} + \cdots + 
\ddot B^{(k)}_{b_2}\right)
\prod_{i=3}^p \sum_{j_i=1}^k \
 \ B_{b_i}^{(j_i)}.
\eqno (4.27)
$$
Using (4.18), (4.22) and (4.23) and repeating computations of (4.25), 
we deduce from (4.27) that 
$$
\ddot B^{(k+1)}_s \le \rt_s {\b \hat \chi^{k-1} \over \r^2}
\left( { 4^3 \a s\over \r}\sum_{p\ge 1}
 {1\over (p-1)!} \left( {8\a\chi \over 1-\hat \chi}
\right)^{p-1} + \left({8\a s\over \r}\right)^2 {4\over 1-\hat \chi}
\sum_{p\ge 2} {1\over (p-2)!} \left( {8\a \chi \over 1-\hat \chi}\right)^{p-2}\right).
$$
Then we can write that
$$
\ddot B^{(k+1)}_s \le \rt_s {\b \hat \chi^{k} \over \r^2}
\left( {1\over 2} + {2\a\chi \over 1-\hat \chi } \right) 
\exp\left\{ {8\a\chi  \over 1-\hat \chi}\right\}.
$$
Taking into account the conditions (4.4), it is easy to see that 
$\ddot B^{(k+1)}_s \le \rt_s {\b \hat \chi^{k} \over \r^2}$. $\Box$

\vskip 0.2cm
According to the general description of the tree-type walks given in sub-section 4.1
and using the second relation of (4.1),
we can write that  
$$
\CZ_{2s}\le n \sum_{h\ge 0}\,  \sum_{\CT_h} \, \sum_{p\ge  0}
{2h\choose p} \, \sum_{l_1\ge 1, \dots, l_p\ge 1}\  
\sum_{a_1+\cdots +a_p=s-h } \prod_{j=1} ^p B^{(l_j)}_{a_j},
$$
where the sums over $l_j$ and $a_i$ are taken over all possible values such that $a_j\ge 2l_j$.  
Using (4.23), we can write that for $\r\ge C$
$$
\CZ_{2s} \le n  
\sum_{h\ge 0} \rt_h \sum_{p\ge 0} {1\over p! } \left( {2s \a\over \r(1-\hat \chi) }\right)^p 
\sum_{a_1+ \cdots+ a_p=s-h} t_{a_1} \cdots t_{a_p} \le 4 n \rt_s\exp\left\{ {8\a\chi \over 1-\hat \chi}\right\}.   
\eqno (4.28)
$$
Then the upper bound (4.2) follows. 

\vskip 0.2cm

To prove relation (4.3), we observe that if  a walk $\CW_{2a}$ belongs to $\ddot \bB^{(k)}_a$, then
it belongs to $ \bB^{(k)}_a$. Therefore $\ddot B^{(k)}_a\le B^{(k)}_a$ and we can write that 
$$
\ddot \CZ_{2s} \le n  \sum_{h\ge 0} \, \sum_{p\ge 0} \ \sum_{\CT_h} V_2^h  \, 
{2h \choose p }
\,  \sum_{l_1\ge 1, \dots l_p\ge 1 } \ 
\sum_{a_1+ \cdots + a_p= s-h} \ 
 p \ddot B_{a_1}^{(l_1)} B_{a_2}^{(l_2)} \cdots B_{a_p}^{(l_p)}.
$$
Then for sufficiently large $\r\ge C$, we get the following upper bound, 
$$
\ddot 
\CZ_{2s} \le n \sum_{h\ge 0} {2\b s\over (1-\hat \chi)\r^2} \rt_h \sum_{p\ge 1} {1\over (p-1)! } \left( {2s \a\over \r(1-\hat \chi) }\right)^{p-1} 
\sum_{a_1+ \cdots+ a_p=s-h} t_{a_1} \cdots t_{a_p} 
$$
$$
\le 
n \rt_s {8 \b \chi\over\r} \exp\left\{ {8\a\chi \over 1-\hat \chi}\right\}
\eqno (4.29)
$$
and relation (4.3) follows. Theorem 4.1 is proved. $\Box$  

\subsection{Proof of Theorem 1.1}

It follows from relations (3.2), (3.3) and Theorem 4.1 that 
$$
M_{2s_n}^{(n,\r_n)} = \dot \CZ_{2s_n}^{(n,\r_n)} + \ddot \CZ_{2s_n}^{(n,\r_n)} + \tilde  \CZ_{2s_n}^{(n,\r_n)}
= \dot \CZ_{2s_n}^{(n,\r_n)}(1+o(1)).
$$
Then  the upper bound (1.6) follows from inequality (4.2). 
Also, it follows from Theorem 4.1 and the first relation of (4.1)  that 
$$
\dot \CZ^{(n,\r_n)}_{2s_n} = n \hat m^{(\r_n)}_{s_n} (1+o(1)),
\eqno (4.30)
$$
where 
$$
\hat m^{(\r)}_s = \sum_{\CW_{2s}\in \bW^{(2,4^\star)  }_{2s}} \pi(\CW_{2s})
\eqno (4.31)
$$
is the total weight of $(2,4^\star)$-walks of $2s$ steps. 

\vskip 0.2cm
{\bf Lemma 4.6.} {\it Denote by $\hat F_\r(z)$ the generating function of the numbers $\hat m^{(\r)}_s$,
$$
\hat F_\r(z) = \sum_{s=0}^\infty \hat m^{(\r)}_s\, z^s.
$$
Then $\hat F_\r(z)$ verifies the following equation,
$$
\hat F_\r(z) = 1+ z\left(\hat F_\r(z)\right)^2 + {z^2 V_4\over \r} \left( {1\over 1-z\hat F_\r(z)}\right)^4.
\eqno (4.32)
$$
}

\vskip 0.2cm
{\it Proof.} Given  a walk $\CW_{2s}\in \bW^{(2,4^\star)}_{2s}$, we will say that it is of $\CM$-type.
Let us consider the first edge
$e_1 = (\vr, \a)$ of the graph $\bar g(\CW_{2s})$ of this walk. If $e_1$ is the $V_2$-edge,
then $\CW_{2s}$  splits into three parts, the sub-walk $(\vr,\a,\vr)$ and two $\CM$-type sub-walks
$\CW_{2a}$ and 
 $\CW_{2b}$, $a+b=s-1$. 

If the edge $e_1$ is the $V_4$-edge, then $\CW_{2s}$ splits in five parts,
the sub-walk $(\vr,\a,\vr, \a,\vr)$ and four sub-walks of $\CS$-type,
$\CW_{2a_i}$, $i=1,2,3,4$ and $a_1+a_2+a_3+a_4=s-4$. We say that  the sub-walk $\CW_{2a}$
is of $\CS$-type if it is
an $\CM$-type sub-walk such that its graph $\bar g(\CW_{2s})$ has the root $\vr\rq{}$ attached by
$V_2$-edges only. 

Let us denote by $S_k$ the total weight of $\CS$-type walks of $2k$ steps.
It is clear that 
$$
S_k = \sum_{l=0}^k\  \sum_{a_1+\dots +a_l=k-l} \ \hat m^{(\r)}_{a_1} \cdots \hat m^{(\r)}_{a_l}.
\eqno (4.33)
$$
It it easy to deduce from (4.33) that 
$$
\sum_{k=0}^\infty S_k z^k = {1\over 1- z\hat F_r(z)}
$$
and then relation (4.32) follows. $\Box$

\vskip 0.2cm
Relation (4.30) combined  with Lemma 4.6 proves
relations (1.8) and (1.9). Theorem 1.1 is proved. 

\vskip 0.5cm
Let us consider the number $m_s^{(\r)}$ and determine its part $[m^{(\r)}_s]_p$ that contains the factor $V_4^p$. 
We can write  that 
$$
[m^{(\r)}_s]_p = \left({V_4\over \r} \right)^p \sum_{h\ge p} {\CT_h\choose p}^\star \ T^{(2p)}_{s-h-p},
\eqno (4.34)
$$ 
where $ {\CT_h\choose p}^\star$ denotes the number of possibilities to choose $p$ edges 
$\hat e_1, \dots, \hat e_p$ in the tree $\CT_h$ 
such that they do not share a common vertex. The factor $T^{(2p)}_{s-h-p}$ counts the number of trees that 
can be attached to the  extremities of the additional edges $\tilde e_i$ 
joined to $\hat e_i$, $i=1,\dots, p$, respectively. Using inequality (4.20), we can deduce from (4.34) that
$$
[m^{(\r)}_s]_p \le  {1\over p!} \left({4s V_4\over \r} \right)^p \sum_{h\ge p} \rt_h \rt_{s-h}\le 
{1\over p!} \left({4s V_4\over \r}\right)^p\, \rt_{s+1}.
\eqno (4.35)
$$
Then the upper bound (1.10) follows.  

\vskip 0.2cm
Regarding the right-hand sides of the upper bounds (3.30) and (3.33), it is easy to see that
$$
{1\over n\rt_s} \tilde \CZ_{2s}^{(n,\r)} = O\left( {s^5\over n}\right), \quad (n,s,r)\to\infty.
$$
This relation together with (4.29) and (4.30) implies (1.12) provided $s=o(n^{1/6})$.





\section{Auxiliary statements}

In this section we collect the auxiliary statements and prove lemmas needed for the proof of Theorems 3.1 and 4.1.

\subsection{Proof of Lemma 3.1}

Let us consider the $q$-vertex $\b$ such that the edge   of the second arrival   $e_2= e(\fa_2) = (\b,\a_2)$
is the minimal  $q$-edge 
over the whole walk $\CW_{2s}$. We denote by $t_2$ the instant of time such that $e_2 = e(t_2)$
and consider the sub-walk $\CW_{[0, t_2-1]}$. The reasonings below concern 
$\CW_{[0, t_2-1]} =\CW^*$ only. 

  If the edge $[\b,\a_2]$ represents the second distinct arrival at $\a_2$ by $\CW^*$, 
then $\a_2$ is the blue $r$-vertex and we are done.
Let us consider the case when $ [\b,\a_2] = E'_1$ is the first distinct arrival at $\a_2$ by $\CW^*$
and denote by $e'_{max} = \max\{ e, \ e\in E_1'(\a_2)\}$. This edge $e'_{max}$ is closed in $\CW^*$ 
by a non-marked edge $f$. We consider two possible orientations of $f$ separately.

 \vskip 0.2cm 
 
 Let us consider first the cases when $f= (\a_2,\b)$. Then $\CW^*$ has to go from $\b$ to $\a_2$
 after $t(f)$ to create the $q$-edge $e(t_2)$. It can arrive at $\a_2$
only with  by a non-marked step $h=(\g,\a_2)$, $\g\neq \b$ that closes the marked
edge $(\a_2,\g) = \hat e$.  Thus, the sub-walk $\CW^*$ 
has to go from $\b$ to $\g$ to perform $h$. If $\CW^*$ arrives at $\g$
by a marked edge $(\d,\g)$, then $\g$ is the blue $r$-vertex because $\d\neq \a_2$. If
$\CW^*$ arrives at $\g$ by a non-marked step $(\d,\g)$, then this step closes 
a marked edge $\{\d,\g\}$. 
If $\{d,\g\} = (\d,\g)$, then $\g$ is the blue $r$-vertex.
If $\{\d,\g\} = (\g,\d)$, then we get the recurrence, where the couple 
$a_2,\g$ is replaced by $\d,\g$. Since $\varkappa_{\CW^*}(\b)= 1$ by$E_1$, then 
this recurrence will be terminated before we come to $\b$ and the $r$-vertex will be specified.

\vskip 0.2cm 
Let us consider the case when $f=(\b,\a_2)$.  To perform this step,
 the sub-walk $\CW^*$ has to go from $\a_2$ to $\b$ before $t(f)$. Assume that it
 arrives at $\b$ by the step $h=(\g,\b)$, $\g\neq \a_2$ that has to be the non-marked one.
 
 Let us consider first the case when $\g\neq \a_1$. The sub-walk has to go from $\a_2$ to $\g$
 and arrive at $\g$ by the step $g=(\d,\g)$. If this step is marked, then $\g$ is the blue $r$-vertex 
 and we are done. If $g$ is non-marked, then it closes the marked edge $\{ \g,\d\}$.
 If $\{\g,\d\}= (\d,\g)$, then $\g$ is the blue $r$-vertex.
 If $\{\g,\d\} = (\g,\d)$, then we get a recurrence. Since $\varkappa_{\CW^*}(\a_2)=1$ by $E_1$, then
 this recurrence will be terminated by a blue $r$-vertex. 
 \vskip 0.2cm
 
 Finally, let us consider the case when $\g=\a_1$ and $h=(\a_1,\b)$.  Then the sub-walk has to go from 
 $\a_2$ to $\a_1$ and arrive it by the step $g=(\g,\a_1)$. If this step is marked, then $\a_1$
 is the blue $r$-vertex. If $g$ is non-marked, then either $\g =\epsilon$ or $\gamma\neq \epsilon$,
 where the edge $(\epsilon,\a_1)\in E_1(\a_1)$.
 
 If $g = \epsilon$, then we get a recurrence with the couple $\a_1, \b$ replaced by $\epsilon, \a_1$. 
 Please note that the fact that $(\epsilon,\a_1)$ generally is not the first arrival at $\a_1$ does not alter
 this recurrence. Then we terminate with the blue $r$-edge.
 
 If $\g\neq\epsilon$, then $g$  closes a marked edge $\{\g,\a_1\}$. If $\{\g,\a_1\}=(\g,\a_1)$, then $\a_1$
 is the blue $r$-vertex. If $\{\g,\a_1\}=(\a_1,\g)$, then we get a recurrence that 
 will terminate before $\a_2$ and  
 the blue $r$-vertex will be specified. Lemma 3.1  is proved. $\Box$


\subsection{Catalan trees and exponential estimates}

In papers  \cite{K01,K3}, the following statement is proved with the help of recurrent relation (2.6).

\vskip 0.2cm
{\bf Lemma 5.1}. {\it Consider the family  of Catalan trees constructed with the help of $s$ edges
and such that the root vertex $\varrho$ has $d$ edges attached to it and denote by $ \rt_{s}^{(d)}$ its cardinality,
$$
\rt_s^{(d)} = 
 \sum_{u_1+\cdots+u_{d-1} +u_d=s-d} \rt_{u_1}\, \rt_{u_2} \cdots \  \rt_{u_{d-1}}\, \rt_{u_d},
$$
where the sum runs over all possible $u_i\ge 0$. 
Then 
the  upper bound 
$$
 \rt_ {s}^{(d)}   \le  e^{- \eta d}  \, \rt_s , \quad \eta = \ln (4/3)
\eqno (5.1) 
$$
is true for any given integers $d$ and $s$ such that $1\le d\le s$. }

\vs
\noindent  {\it Remark.}  We can say that 
$\rt_s^{(d)}$ represents the number of Catalan trees such that their root vertex $\vr$ has the exit sub-cluster
of cardinality $d$. It is not hard to deduce from (2.6) that the numbers $\{\rt_s^{(d)}, 1\le d\le s\}$ 
verify the following recurrent relation \cite{K01},
$$
\rt^{(d)}_s = \rt^{(d-1)}_{s} - \rt_{s-1}^{(d-2)}\, , \quad 3\le d\le s
\eqno (5.2) 
$$
with the initial values $\rt^{(1)}_s = \rt_{s-1}$, $s\ge 1$ and $\rt^{(2)}_{s} = \rt_{s-1}$, $s\ge 2$. 
Denoting $\rt^{(d)}_s = T^{(d)}_{s-d}$ (4.17) and changing variables by $k=s-d$, $p=d$, we deduce from (5.1) and
an elementary consequence of (1.7) $\rt_{s+1} \le 4\rt_s, s\ge 0$ the following  inequalities,
$$
T_k^{(p)} \le e^{-\eta p}\, \rt_{k+p} \le 4^p\, \rt_k.
$$
This proves the upper bound (4.20).

\vskip 0.2cm
Regarding the numbers $R_s= T^{(3)}_{s-2} $ (4.17), it is easy to see that 
$$
\sum_{s=2}^\infty R_s z^s = z^2f^3(z).
$$
It follows from (1.11) that $zf^2(z) = f(z)-1$. Using this equality several times, we obtain that 
$$
R_s = 3 {(2s-2)!\over (s-2)! (s+1)!}, \quad s\ge 2.
$$
This proves the last relation of (4.17). 
\vskip 0.3cm

Let us denote by $\CN^{(1,2)}_s $ the number of tree-type walks of $2s$ steps such that their graph
contains one $V_4$-edge $(\a,\b)$ and remaining $s-2$ edges are $V_2$-edges.
Then it is not hard to see that (cf. (4.17))
$$
\CN^{(1,2)}_s = \sum_{a+ b_1+_2+b_3 = s-2} (2a+1) \rt_a \, \rt_{b_1} \, \rt_{b_2} \, \rt_{b_3}, \quad s\ge 2,
$$
where the factor $(2a+1)$ gives the number of choice of the root $\r$ in the sub-tree $\CT_a$
attached to the vertex $\a$ while the remaining three sub-trees $\CT_{b_i}$ are attached to the
vertex $\b$. Then the generating function 
$$
\Phi^{(1,2)}(z) = \sum_{s=2}^\infty \CN^{(1,2)}_s z^s , \quad \CN^{(1,2)}_2=1
$$
is given by expression 
$$
\Phi^{(1,2)}(z) = 2z^3 f'(z) f^3(z) + z^2 f^4(z).
$$
Using (1.11), one can show that (see \cite{K3} for details)
$$
\CN^{(1,2)}_s = {(2s)!\over (s-2)!\, (s+2)!} = s \rt_s \left( 1- {3\over s+2}\right), \quad s\ge 2.
$$
More generally, denoting by $\CN_s^{(1,m)}$ the number of even closed  walks
$\CW_{2s}$ such that their graphs contain one edge of total multiplicity $2m$ 
and all other edges of multiplicity $2$, one can prove that relation 
$$
\CN_s^{(1,m)} = {(2s)!\over (s-m)! \, (s+m)!}, \quad s\ge m\ge 1
\eqno (5.3)
$$
is true. The  proof of (5.3)  can be obtained with the help of the 
following reasoning similar to that used in \cite{K01} to deduce relation (5.2).

Let us introduce variables 
$$
D^{(m)}_k= \sum_{a+b_1 + \dots + b_{2m-1} = k-m} \ (2a+1) \rt_a \rt_{b_1} \cdots \rt_{2m-1}, \quad k\ge m\ge 1,
$$
such that $D^{(l)}_k= 0$ for all $0\le l < k$   
and 
$$
E^{(m)}_k = \sum_{a+b_1 + \dots + b_{2m-1} = k-m} \ (2a+1) \rt_a \rt_{b_1} \cdots \rt_{2m}, \quad k\ge m\ge 1,
$$
such that $E^{(l)}_k = 0$ for all $0\le l< k$.
It is not hard to see that $D^{(m)}_s = \CN^{(1,m)}_s$. Indeed, regarding a vertex $\a$, we attach to it $m$ blue edges and get 
the sub-cluster $\D_m$.
Then we attach to $\a$ a green Catalan tree $\CT_a$ and determine the root $\vr$ by choosing one of $2a+1$ instants
of time of the chronological run over $\CT_a$.  
To we attach $2m-1$ red Catalan trees $\CT_{b_i}$ at the remaining $2m-1$ instants of time of the chronological run
over the sub-cluster $\D_m$ of blue edges.

According to  these definitions, we get that 
$$
D^{(1)}_k = \sum_{a+b=k-1} (2a+1) \rt_a \rt_b, \quad k\ge 1
$$
and $D^{(1)}_0= 0$. Then the generating function  
$\CD^{(1)}(z) = \sum_{k\ge 0} D^{(1)}_k z^k$
is such that 
$$
\CD^{(1)}(z) = 2z^2 f'(z) f(z) + z(f(z))^2 = zf'(z)
$$
and therefore 
$$
D^{(1)}_k = {(2k)!\over (k-1)!\, (k+1)!},\quad  k\ge 1.
$$ 
In this computation,  we have used the identity
$2zf'(z) f(z) = f'(z) - (f(z))^2$ that follows from  (1.11). 
We can also wright that $D^{(1)}_k= k \rt_k$. This relation is obvious because
the number
 $D^{(1)}_k$ by its definition enumerates
the ensemble of Catalan trees with one marked edge colored in blue.

\vskip 0.2cm
It follows from the definition of $E^{(m)}_k$   that the generating function 
$\CE^{(1)}(z) = \sum_{k\ge 0} E^{(1)}_k z^k $ is given by the formulas
$$
\CE^{(1)}(z) = 2z^2 f'(z) (f(z))^2 + z (f(z))^3 = {f'(z)\over 2} - {f(z)-1\over 2z}
$$
and therefore
$$
E^{(1)}_k = {k\over 2} \rt_{k+1} , \quad k\ge 0.
$$

Using the fundamental recurrence (2.6), we can write that 
$$
D^{(m)}_k = \sum_{a+b_1 + \dots + b_{2m-3} +c = k-m} \ (2a+1) \rt_a \rt_{b_1} \cdots \rt_{2m-3} \sum_{b_{2m-2} + b_{2m-1} = c} 
\rt_{2m-2} \, \rt_{2m-1}
$$
$$
=  \sum_{a+b_1 + \dots + b_{2m-3} +c = k-m} \ (2a+1) \rt_a \rt_{b_1} \cdots \rt_{2m-3} \rt_{c+1}.
$$
Then we get the following  equality,
$$
D^{(m)}_k =  E^{(m-1)}_k - D^{(m-1)}_k, \quad k\ge m\ge 2.
$$
Similar computation shows that 
$$
E^{(m)}_k = D^{(m)}_{k+1} - E^{(m-1)}_k, \quad k\ge m\ge 2.
$$
Using these recurrent relations together with the initial expressions given by $D^{(1)}_k$ and $E^{(1)}_k$,
one can easily check that 
$$
D^{(m)}_k = {(2k)!\over (k-m)! \, (k+m)!} \quad {\hbox{and }} \quad 
E^{(m)}_k = { (2k+1)!\over (k+1-m)!\, (k+m)!}
$$
for all $k\ge m\ge 1$. This proves relation (5.3).
\vskip 0.2cm

Using the same reasoning as above, it is not hard to see  that relation (5.2) 
implies that 
$$
\rt_s^{(d)} = 
 \begin{cases}
 (2m-1) \rt_{s-m}\, \prod_{i=1}^{m-1} {\displaystyle  s+1-m-i\over \displaystyle  s+1-i}  , & \text{if  $d=2m-1,\,  m\ge 1$}, \\
m\rt_{s-m}\, \prod_{i=1}^{m-1} { \displaystyle  s-m-i\over \displaystyle   s+1-i}, & \text {if  $d= 2m, \, m\ge 0$.}
\end{cases}
\eqno (5.4)
$$

\vskip 0.2cm
Let us denote by $\CN^{(2,2)}_s$ the number of tree-type walks that contain two $V_4$-edges while the remaining ones are 
the $V_2$-edges. One can show that 
the generating function $\Phi^{(2,2)}_s$ is given by the following expression,
$$
 \Phi^{(2,2)}(z) = {z^4\over 2} f''(z) f^4(z) + 3 z^4 f'(z) f^6(z).
 $$
 Then 
 $$
 \CN^{(2,2)}_s =  {\rt_s}   { k^4 + 8k^3 + 39k^2 +12\over 2(k+2)(k+3)}  - 4^k + 
 3 {(2k)!\over (k-4)!\, (k+4)!}.
 $$
 Although the last expression  is not so compact as those of  (5.3), we can easily deduce from 
 them 
 that 
 $$
 \lim_{s\to \infty} {1\over s\rt_s} \CN^{(1,2)}_s = 1\quad {\hbox{and }}\quad 
 \lim_{s\to \infty} {1\over s^2\rt_s} \CN^{(2,2)}_s = {1\over 2}.
 $$
 These relations agree with the upper bounds (4.35) in the cases of $p=1$ and $p=2$. Moreover,
 one can put forward a conjecture  that 
 $$
 \CN^{(p,2)}_s = {s^p\over p!} \, \rt_s (1+o(1)), \quad s\to\infty. 
 $$
This allows one  to expect that the following lower bound holds (cf. (1.10)),
$$
\liminf_{(n,s,\r)\to\infty }\,  {1\over  \rt_s} 
\hat m^{(\r)}_s \ge e^{  {\chi V_4}}. 
$$
Finally, let us note that a part of $\CN^{(2,2)}_s$ given by 
$$
\ddot \CN^{(2,2)}_s = 4{(2s)!\over (s-4)!\, (s+4)!}
$$
represents the number of $(2,4)$ walks of $2s$ steps such that have two $V_4$-edge with common vertex. It is easy to see that 
$\ddot \CN^{(2,2)}_s = s\rt_s(1+o(1))$ and therefore these walks do not contribute to the asymptotic expression for $M^{(\r)}_{2s}$.
This is in complete accordance with the definition of the numbers $\hat m_s^{(\r)}$ as the total weight of the tree-type $(2,4^\star)$-walks.
The terms $\ddot \CN^{(2,2)}_s$ and $\CN^{(1,3)}_s$ provide the leading contribution to the asymptotic expansion (1.13).


\subsection{D-lemma}

In the present subsection we prove Lemma 3.4.
Let us introduce an auxiliary   collection of variables 
$$
  \CH =  (\bar m_I,  (\bar \L, \Psi, \bar f'')_J, (\Phi, \bar f')_K)
$$
that represents a part of parameters $\CP_R (\bar x, \bar y, \bar z, \CH)$ (3.17)
and consider its numerical  realization $\la \CH\ra$. 
Then relation (3.24) can be rewritten in the following form, 
$$
\vert  \sqcup_{\la \CH\ra}  \bW_{2s}(D, \la \breve v\uplus \CG^{(c)}\ra_s, \la \CH\ra,  \U)\vert \le 
2^{\vert \bar q\vert } \, D^{\vert \bar p\vert } \, \left( e^\eta h_0^2\right)^{I+J+K} \, e^{ - \eta D + eD/h_0} \, \rt_s,
\eqno (5.5)
$$
where the disjoint union is taken over the set of all possible realizations  $\bH =\{ \la \CH\ra\}$.
By the construction, the values of   $\la (\bar x, \bar y, \bar z)\ra_s$
are determined by the realization of the color diagram $\la \CG^{(c)}\ra_s$. As we will see below, 
the set $\left( \la\bar x, \bar y, \bar z\ra_s,  \la \CH\ra\right)$ uniquely determine
the nest cells $\check \u_1, \dots, \check \u_R$ in the underlying trees $\CT(\CW_{2s})$ where the clusters
$\check \D_1, \dots, \check \D_R$ are attached. 
Then we can apply inequalities of the form (5.3) to get the upper bound  of the set 
of underlying trees that is is exponential with respect to the sum $\sum_{i=1}^R d_i$, $d_i = \vert \check \D_i\vert$.

We prove (5.5) by recurrence with respect to $N=I+J+K$.

\subsubsection{The case of $ R=1$}

 If  the total number of cells at $\breve \b$ is equal to one, $R=1$, 
then either $\CP_1 = x_1$ or $\CP_1= z_1$  and the set of variables $\CH$ is empty. For simplicity, we  consider  the 
 former case such that   $\la x_1 \ra_s= \t_1\ge 1$.
 Regarding a walk $\CW_{2s}$ from the left-hand side of (5.5)
and the corresponding tree $\CT_s = \CT(\CW_{2s})$, we observe that its vertex 
$\check  \u$ such that $\check  \u = \fR(\xi_{\t_1})$ is attached by a sub-cluster $\check \D_1$ of $d_1$ edges.
It is easy to construct the corresponding family of trees $\bT_s(\t_1,d_1)$ with the help of  the following procedure. 

Let us take a root vertex $\fb_0$ and attach to it a linear branch $\CB_l$ that consists of  $l$ edges
and $l+1$ vertices $\fb_0, \fb_1, \dots , \fb_l$. 
Regarding the set of vertices  $\{\fb_0, \dots, \fb_{l-1}\}$, we attach to them 
  the sub-trees $\CT_{a_1}, \dots,
\CT_{a_l}$ with given $\bar a= (a_1,\dots, a_l)$ 
such that $\vert \bar a\vert = a_1 + \dots + a_l = \t_1-l$. We 
do this in the way that the sub-trees grow to the left of the branch $\fB_l$
with respect to the ascending chronological run over $\fB_{l-1}$ from $\fb_0$ to $\fb_{l-1}$.

We attach to  the vertex  $\check    \u=\fb_l$ the sub-cluster $\check \D_1$ of $D=d$ edges. 
Using $d$ vertices of $\check \D_1$, we attach to them  $d$ sub-trees
$\CT_{b_1}, \dots , \CT_{b_{d}}$. 
 Regarding the vertices $\fb_{l_1}, \dots, \fb_0$ as  the descending part of the chronological
run over $\CB_{l-1}$, we construct on these vertices the sub-trees
$\CT_{c_1} ,\dots , \CT_{c_l}$, where $b_i$ and $c_j$ 
are such that $\sum_{i=1}^d b_i+ \sum_{j=1}^l c_j = s-\t_1 - d$.

Then we can write that 
$$
\RT_s(\t_1, d)= \vert  \bT_s(\t_1,d)\vert =
\sum_{l=1}^{\t_1} \ \  \sum_{
\stackrel{\bar a:}{\vert \bar a\vert=\t_1-l}
} \rt_{a_1} 
\cdots
\rt_{a_l} \
\sum_{ 
\stackrel{\bar b, \bar c:}
{\vert \bar b \vert +\vert \bar c\vert = s-\t_1-d} }
\rt_{b_1} \cdots \rt_{b_{d}}\cdot \rt_{c_1} \cdots \rt_{c_l}.
\eqno (5.6)
$$
It follows from Lemma 5.1 that 
$$
\sum_{\bar b: \ \vert \bar b \vert = m} \ \rt_{b_1}\cdots \rt_{b_d} \le \ e^{-\eta d} \, \rt_{m+d}.
\eqno (5.7)
$$
Then we can  deduce from  (5.6) the following inequality,
$$
e^{\eta d }\, \RT_s(\t_1, d)\le \
\sum_{l=1}^{\t_1} \ \ 
\sum_{
\stackrel{\bar a, b, \bar c:\ \vert \bar a\vert, b, \vert \bar c\vert\ge 0 }{\vert \bar a\vert + b + \vert \bar c \vert = s-l}} 
\rt(\bar a) \cdot  \rt_b \cdot  \rt(\bar c) 
=    \rt_s,
\eqno (5.8)
$$
where we denoted $\rt(\bar a) = \rt_{a_1}\cdots \rt_{a_l}$.
The last equality follows from the observation that the sum in the central part  of (5.8)
represents the cardinality of the set of Catalan trees $\CT_s$ that have  the vertex $\check \u$  seen  
 at the instant $\xi_{\t_1}$ of the chronological run $\CR\{\CT\}$  colored in white, the others  being the black ones.
Clearly, given $\CT_s$, there exists 
only one such vertex and therefore the cardinality of this family is equal to $\rt_s$.

Combining inequality (5.8) with the filtration estimate (3.9a), we get the needed upper bound (3.9b),
$$
\vert \bW_{2s} (D, \breve v(x_1)\uplus \la \CG^{(c)}(\bar \nu,\bar p, \bar q)\ra_s^{(b)}, \U)\vert 
\le 2^{\vert \bar q\vert } \, D^{\vert \bar p\vert} \, e^{-\eta D} \, \rt_s.
\eqno (5.9)
$$

\subsubsection{The cases of $R=2, N=2$}

a) Let $\CP_2$ be such that  $\bar x = (x_1,x_2)$ and $\la \bar x\ra_s=(\t_1, \t_2), 1\le \t_1<\t_2$. 
Then  $\CH$  is empty.
The mirror cells are not present at $\bb$ and
therefore the tree $\CT_s =\CT(\CW_{2s})$ is such that the vertices 
$\check \u_1$ and $\check  \u_2$
lie on  two different branches $\CB_1$ and $\CB_2$, respectively. 
Let us describe the construction of the corresponding subset of trees and estimate its cardinality.

We start with the first branch $\CB_1 = \CB_{l_1}$ that contains $l_1$ edges and $l_+1$ vertices
$\fb_0, \fb_1, \dots, \fb_{l_1}$
and construct on its first $l_1$ vertices the sub-trees $\CT_1, \dots, \CT_{a_{l_1}}$
such that $\vert \bar a\vert = \t_1-l_1$. 
Then we attach to $\fb_{l_1}=\check \u_1$ the sub-cluster $\check \D_1$ of $d_1$ edges
and join to each of $d_1$ vertices the sub-trees 
$\CT_{b^{(1)}_1},\dots, \CT_{b^{(1)}_{d_1}}$
such that $\vert \bar b^{(1)}\vert = m_1\ge 0$. 

Performing $\l$ steps down along  the descending part of $\CB_1$, 
we stop at the vertex 
$\fb_{l_1-\l}$, $1\le \l\le l_1$
and attach to it the second branch $\CB_2$ with vertices $\fc_0= \fb_{l_1-\l}, \fc_{1},\dots, \fc_{l_2}$.

Regarding vertices $\fb_{l_1-1}, \dots, \fb_{l_1-\l}, \fc_1,\dots \fc_{l_2-1}$, 
we construct attach to them the sub-trees 
$\CT_{c^{(1)}_j}$ such that 
$\vert \bar c^{(1)}\vert =c_1^{(1)}+\dots c_\l^{(1)} +c_{\l+1}^{(1)}+
\dots+c_{\l+l_2-1}^{(1)}= \t_2-\t_1-l_1-l_2-m_1$.

We construct the second sub-cluster $\check \Delta_{d_2}$ with $\vert \check \Delta _{d_2}\vert = d_2$
on the vertex $\fc_{l_2}=\check \u_2$ and construct on the $d_2$ vertices obtained 
the sub-trees $\CT_{b^{(2)}_1}, \dots, \CT_{b^{(2)}_{d_2}}$. Finally, we 
 join the sub-trees $\CT_{c_1^{(2)}}, \dots, \CT_{c_{l_1+l_2-\l}^{(2)}}$ to the  vertices 
$\{\fc_{l_2-1}, \dots, \fc_1, \fb_{l_1-\l}, \dots, \fb_0\}$. 
Then we can write that 
$$
\vert \bT_s(\t_1, d_1;\t_2, d_2)\vert 
= \sum_{l_1=1}^{\t_1}  \sum_{\l=1}^{l_1} \sum_{l_2=1}^{\t_2-l_1-d_1}
\sum_{\bar a :\,  \vert \bar a\vert = \t_1-l_1}\rt(\bar a) \  
\sum_{m_1=0}^{\t_2-\t_1-l_1}\ \sum_{\bar b^{(1)}:\,  \vert \bar b^{(1)} \vert = m_1}
\, \rt(\bar b^{(1)})
$$
$$
\times\  \sum_{\bar c^{(1)}: \, \vert \bar c^{(1)}\vert =0}
^{ \t_2-\t_1-d_1-m_1-l_2} \, \rt(\bar c^{(1)})
\ \
\sum_{m_2=0}^{s-\t_2-d_2} \ \  \sum_{\bar b^{(2)} : \ \vert \bar b^{(2)} \vert = m_2} \ 
\rt(\bar b^{(2)}) 
\ \ \sum_{\bar c^{(2)}: \, 
\vert \bar c^{(2)}\vert=  s-\t_2-d_2- m_2} \,
\rt(\bar c^{(2)}).
\eqno (5.10)
$$

We apply (5.7)
two times with respect to the sub-trees  with the exit sub-clusters $\check \D_1$ and $\check \D_2$ and get the estimate
$$
\sum_{\bar b^{(1)}: \ \vert \bar b^{(1)} \vert = m_1} \ \rt(\bar b^{(1)})
\  \ 
\sum_{\bar b^{(2)} : \ \vert \bar b^{(2)} \vert = m_2} \ \rt(\bar b^{(2)})
\le \ e^{-\eta (d_1+d_2)} \, \rt_{m_1+d_1} \, \rt_{m_2+d_2}
\eqno (5.11)
$$
The same reasoning as before show that 
 inequality of the form (5.8) 
with $d= d_1+d_2$, where   $\rt_s$ represents the number of Catalan trees
such that  the vertices seen at the instants $\xi_{\t_1}$ and $\xi_{\t_2}$ are colored in white. 
Using (3.23), it is easy to complete the proof of (5.5).

\vskip 0.5cm 
b) Let us consider the case of two cells $\CP_2\ = (x_1, (y_1,\L))$
such that  $\la (x_1, x_2)\ra_s = (\t_1, \t_2)$ are  given as well as a particular value $\la \L\ra= \l\rq{}$. Then
$\bb$ is attributed by 
one proper cell and one imported cell. 
We first study  the case when  $y_1$ does not fill the edge-box attached to  a red or to a green vertex of $\CG^{(c)}$. 
There is no mirror cells at $\bb$ and therefore the vertices 
$\check \u_1 = \fR(\xi_{\t_1}) $ and $\check  \u_2 = \fR(\xi_{\t_2})$ are situated on different branches
of $\CT_s$. 
Let us briefly describe the construction  of the corresponding tree $\CT_s$ that is very similar to that we performed above.

Taking the root vertex $\fb_0$, we draw  a branch $\CB_1$ with the help of $l_1$ edges. 
Starting from the extreme vertex $\fb_{l_1}$,  we descend by  $\l_1$ steps till the vertex 
$\fb_{l_1-\l_1}$ and attach to it the second branch $\fB_2$ of $l_2$ edges. We attach to the vertex $\check \u_1$
the sub-cluster $\D_1$ of $d_1$ edges. 
We denote the skeleton obtained 
by $K(l_1,d_1,\l_1;l_2)$. Regarding the vertices of $K$, we construct on them the subtrees
$\CT(\bar a)$, $\CT(\bar m_1)$, $\CT(\bar b)$ and $\CT(\bar c^{(1)})$ with 
properly chosen values. In this construction,
we do not use the vertices of the descending part of $K$ from the vertex $\check \u_2$ to $\fb_0$.
We denote the sub-tree obtained by $\grave \CT = \CT(\bar a, \bar \b, \bar c^{(1)}, \bar m_1; K)$.

\vskip 0.2cm
Now let us consider a sub-walk $\CW_{[0, \xi_{\t_2-1}]}(\grave \CT)=\grave \CW$ performed according to the
rules of $\la \CG^{(c)}\ra_s$ and $\Upsilon$. The vertex $\breve \b$ is completely determined by the run of 
$\grave \CW$ as well as the vertex $\CW(\xi_{\t_2})= \gamma$. 
Therefore the path $\CL$ from $\g$ to $\breve \b$ by non-marked steps according to $\U$, if it exists,
is completely determined as well as its length $\vert \CL\vert = \l_2$. This produces
the indicator function $I_{\grave \CW}(\lambda\rq{})$ that is equal to   1 if $\l\rq{}=\l_2$ and zero otherwise.

\vskip 0.2cm

With  $\l_2$ determined, we descend from $\check \u_2$ to $\fb_0$ of $K$ by $\l_2$ steps
and attach to the vertex obtained the sub-cluster $\D_2$ of $d_2$ edges. Then the family of 
sub-trees $\acute \CT(d_2) = \CT_{s-\t_2}(\bar c^{(2)}, \bar m_2)(d_2)$ with the help of the remaining edges is constructed.

\vskip 0.2cm

Regrading the sum over all values of $\la \CH\ra=\l\rq{}$, we see that the cardinality of the set of trees 
obtained is given by the following expression (cf. (5.10)),
$$
\sum_{\l\rq{}} \ \sum_{l_1, \l_1, l_2} \ 
\sum _{K(l_1, d_1,\l_1;l_2)} \ \sum_{\grave \CT} I_{\grave \CW}(\l\rq{}) \cdot \sum_{\acute \CT} 1.
\eqno (5.12)
$$
Taking into account that 
$$
 \sum_{\acute \CT_{s-\t_2}(\bar c^{(2)}, \bar m_2)(d_2)} 1 \le e^{-\eta d_2}\ 
\sum_{\acute \CT_{s-\t_2}(\bar c^{(2)}, \bar m_2)} 1
\eqno (5.13)
$$
and that $\sum_{\l\rq{}} I_{\grave \CW}(\lambda\rq{})=1$, we conclude that the right-hand side of 
(5.12) is bounded by the sum
$$
e^{-\eta d_2} \  \sum_{l_1, \l_1, l_2} \ 
\sum _{K(l_1, d_1,\l_1;l_2)} \ \sum_{\grave \CT} \ \sum_{\acute \CT} \, 1
\le e^{-\eta (d_1+d_2)} \ \rt_s.
$$
To get the last inequality, we have used the same reasoning as that of (5.6), (5.7) and (5.8).
Then (5.5) follows.

\vskip 0.5cm
c)  Let us consider the case of $\la \CP_2\ra =(\t_1,  (\t_2, \l\rq{}))$ such that 
the variable $ y_1 $ is attributed to
 the  second arrival at the green vertex $\hat v$ of $\CG^{(c)}$. The case when it is attributed to the red edge-box
is similar and we do not discuss it here.

Let us denote by $q\rq{}$ and $p\rq{}$ the number of red and green vertices that lie to the left of 
$ \hat v$ and by $q\rq{}\rq{}$ and $p\rq{}\rq{}-1$ the number of red and green vertices to the right of $\hat v$.
We construct the skeleton $K$ and the tree $\grave \CT$ as it is described above.
Then we perform the run of the sub-walk $\grave \CW= \CW^{(\grave \CT)}_{[0, t_2-1]}$, $t_2 = \xi_{\t_2}$
following the prescriptions of $\la \breve v \uplus \CG^{(c)}\ra_s^{(b)}$ and the rule $\U$.
The vertex $\a = \grave \CW(t_2-1)$ being determined, 
the exit cluster $\D(\a)= \{\g_1, \dots, \g_m\}$ is also uniquely determined. 

At the instant of time $\grave \xi_{\t_2}$ the walk has to choose a vertex $\gamma\rq{}$ from $\D(\a)$ such that 
$\g\rq{}$ is situated on the distance of $\la \L\ra=\l\rq{}$ non-marked steps from $\bb$. 
Therefore the indicator function of (5.12) $I_{\grave \CW}(\l\rq{})$ is replaced by 
$I_{\grave \CW}^{(\g_1,\dots, \g_m)}(\l\rq{})$ that is non-zero only in the case when $\l\rq{}$ takes one of the
values that correspond to the length of one of  the paths of non-marked edges from $\g_i$
to $\breve \b$, if such paths exist. Thus,
$$
\sum_{\l\rq{}} I_{\grave \CW}^{(\g_1,\dots, \g_m)}(\l\rq{})
\le m = \vert \D(\a)\vert \le D.
\eqno (5.14)
$$

Denoting $\bW_{2s}^* =  \bW_{2s}(d_1,d_2; \la \breve v\uplus \CG^{(c)}\ra_s, \la \CH\ra,  \U)$,
we can write the following equality
$$
\bW^*_{2s} = \sqcup_{\grave \CT}  \left\{ \CW_{[0, \xi_{\t_2-1}]}\right\}
\otimes \left\{ \la \CW( \xi_{\t_2})\ra_{\l\rq{}}\right\}
\otimes \sqcup_{\acute\CT}  \left\{ \CW_{[\xi_{\t_2}+1, 2s]}^{(\acute \CT)}\right\},
$$
where the curly  brackets denote  the families of realizations of corresponding sub-walks and 
$ \la \CW( \xi_{\t_2})\ra_{\l\rq{}}$ denotes the set of  possible values $\g\rq{}$. 
Then 
$$
\sum_{\l\rq{}} \vert \bW^*_{2s}\vert \le 
\prod_{\grave  \CT} \# \left\{ \CW_{[0,  \xi_{\t_2-1}]}\right\} \times
\sum_{\l\rq{}} \# \left\{ \la \CW( \xi_{\t_2})\ra_{\\l\rq{}}\right\}
\times e^{-\eta d_2} \, 2^{q''} \, D^{p''-1}  \ \#\left\{\acute \CT\right\}, 
$$
where we have used inequality (5.13).

It follows from (5.14) that 
$$
\sum_{\l\rq{}}\# \left\{ \la \CW( \xi_{\t_2})\ra_{\\l\rq{}}\right\}\le D.
$$
Using this  inequality and estimates 
$$
\# \left\{ \CW_{[0, \grave \xi_{\t_2-1}]}\right\}\le \, 2^{q'} \, D^{p'} 
$$
and 
$$
\prod_{\grave  \CT} 1 \cdot \#\left\{\acute \CT\right\}  \le e^{-\eta d_1} \, \rt_s,
\eqno (5.15)
$$
we conclude that  
$$
\sum_{\l\rq{}} \vert \bW^*_{2s}\vert\le  2^{\vert \bar q\vert } \, D^{\vert \bar p\vert }\,  e^{-\eta (d_1+d_2)} \rt_s.
$$
Remembering (3.23),  it is easy to show that (5.6) is true in the case under consideration.

\subsubsection{The cases of $N=2, R\ge 3$}

Let us consider $\CP_3$ such that  the first two  cells at $\bb$ are given by  the instants $\la (x_1,x_2)\ra = (\t_1,\t_2)$ 
while the third one is represented by 
the mirror cell. 
The presence of this  mirror, $m_2 =1$ means that 
in the Dyck-type part of the walk, and in the corresponding tree the vertices $\check \u_1 = \fR(\xi_{\t_1})$ and 
$\check \u_2 = \fR(\xi_{\t_2})$ lie on the same branch of edges that starts at the root vertex $\fb_0$.
Therefore the tree $\CT$ is of the following structure: 
we choose a length $l_1$ and construct the branch $\CB_1$ of $l_1 $ edges that starts by $\fb_0$ and ends by 
$\check \u_1=\fb_{l_1}$. 
Then we attach to $\check \u_1$ another linear branch $\CB_2$ of $l_2$ edges that ends by $\check \u_2$. 
We attach the exit sub-cluster $\check \D_1$ to $\u_1$ at the instant  $l_1$ of the chronological run $\fR(\CB_1\uplus \CB_2)$
and the sub-cluster $\check \D_3$ to the vertex $\u_1$ at the instant $l_1 + 2l_2+1$ of $\fR(\CB_1\uplus \CB_2)$.

Then we attach 
 the sub-cluster $\check \D_2$ at the vertex $\check \u_2$. Regarding $2(l_1+l_2)+1 + d_1+d_2+d_3$ vertices of the obtained 
 skeleton, we attach to them  the sub-trees of the total number of edges $s-(l_1+l_2+d_1+d+2+d_3)$. 
Using three times inequalities of the from (5.7) we easily get exponential estimates of (5.6) in this case.
 \vskip 0.5cm
 
To complete the study of the initial step of the proof of  Lemma 5.1, we  consider the case of
numerous imported cells of the form  
 $\CP_{R} = (z_1, (y_1, \L, \psi_1, \dots, \psi_{f}))$, where $f=f''_1$ and $R=3+f$.
 We assume for simplicity that $\la (z_1,y_1)\ra= (\t_1,\t_2)$ and that $\t_1<\t_2$.
The reasoning presented below can be applied without any changes to the case of imported cells generated by the local BTS 
instants $\la (z_1,z_2) \ra = (\t_1,\t_2)$. 
Let us point out  that in this situation  either $f=0$ or $f=1$ (see inequality (3.22)).
However, we include  into considerations the general case of greater values of $f$.
Another remark is that  we can  ignore the presence of the proper cell $\la z_1\ra =\t_1$ with the exit sub-cluster $\check \D_1$ 
at $\bb$ and consider the imported cells and corresponding exit sub-clusters only.
We also assume for simplicity that $y_2$ is  attributed to a blue $r$-vertex $\hat v $ of $\CG^{(c)}$.

To get a realization of $\la \CH\ra$, we take an integer  $f$ and then attribute numerical  values 
to the variables $\L, \psi_1, \dots, \psi_{f}$ given by $\l\rq{}, \psi_1\rq{}, \dots, \psi_{f}\rq{}$. 
Let us take a tree $\grave \CT = \grave \CT_{\t_2}$ and consider a part
of the chronological run over it $\fR_{[0, t'-1]} $ with $t'= \xi_{\t_2}$. Following this run, we construct
a sub-walk $\CW_{[0,t'-1]}$ according to the rules prescribed by $\la \breve v \uplus \CG^{(c)}\ra_s^{(b)}$ and $\U$. 
At the instant of time $t'$, the walk has to join a vertex $\g$ of $g(\CW_{[0,t'-1]})$ prescribed by the values of marked instants 
of the edge-boxes attached to $\hat v$. This vertex $\g$ is uniquely determined
and therefore we are able to conclude whether the set of numerical data $f,  (\l\rq{}, \psi_1\rq{},  \dots, \psi_f\rq{})$ 
is compatible  with $\CW_{[0,t'-1]}$ or not. We mean that it becomes clear whether 
there exists a path from $\g$ to $\bb$ of $\l\rq{}$ non-marked steps that the walk can perform
according to the rules $\U$ or not. The same concern $f$ consecutive returns to $\bb$
with the help of $\psi_l\rq{}$ non-marked steps. 

The $f+1$ nest cells are uniquely determined in $\grave \CT_{\t_2}$ and the exit sub-clusters
of the total cardinality $\check D= D- (f+1)$ are to be distributed to these nest cells. Let us denote by 
 $\bar d_{f+1}$ this distribution. 
 We also denote by $\bT_s(\grave \CT_{\t_2} \uplus \{ \check \D_1, \dots, \check \D_{f+1}\})$
 a collection of Catalan trees constructed over the base tree $\grave \CT$ with the exit sub-clusters $\check \D_j$ attached. 
 
 Using (5.8) and (5.11) several times,  
one can easily prove the exponential estimate for the number of trees 
$$
\vert \bT_s( \grave \CT_{\t_2} \uplus \{ \check \D_1, \dots, \check \D_{f+1}\})\vert \le e^{-\eta \check D} \vert \bT_s(\CT_{\t_2})\vert .
$$

By changing  somehow the point of view,
 we can say  that given $\la \breve v \uplus \CG^{(c)}\ra_s^{(b)}$ and $\U$,
 the set of all possible values of $f$ and $ \la ( \L, \psi_1,  \dots, \psi_f)\ra$  is filtered by the run of the walk
$ \CW_{[0,t'-1]}$. The values $f$ and $\check D = D - (f+1)$ depend on the realization of $\CW_{[0,t'-1]}$.
With the help of the filtration principle, we get the following inequality, 
 $$
 \vert\sqcup_{\la \CH\ra} 
  \bW_{2s}^{(\grave \CT)}(D, \la \breve v \uplus \CG^{(c)}\ra_s^{(b)}, \la \CH\ra,  \U) \vert 
 $$
$$ \le 
 2^{\vert \bar q\vert} \, D^{\vert \bar p\vert }\, 
  \sup_{f} \left\{ e^{\eta (f+1)} 
  {\check D + f \choose f} \right\}
 \, e^{-\eta  D}\ \sum_{\grave \CT} \vert \bT_s(\CT_{\t_2})\vert,
 \eqno (5.16)
$$
where the superscript $\grave \CT$ means that the walks have this tree as the first part of the underlying trees.
Taking into account the upper bound $f \le K =1$ (see (3.22)), we 
can apply to the right-hand side of (5.16) relations (3.21) and  (3.23)
and write that 
$$
 \sup_{f} \left\{ e^{\eta (f+1)} 
  {\check D + f \choose f} \right\}
\le 
e^{2\eta}\,  h_0^2 \, e^{eD/h_0}.
\eqno(5.17)
$$
Repeating the reasoning of (5.8), we  get from (5.16) and (5.17) the upper bound  (5.6).

\subsubsection{The general step of recurrence}

The general step of recurrent estimate of (5.6) is to show that if this estimate is true for $N=I+J+K$,
then it is true in the case of $N' = N+1$, where $N' = I'+J'+K'$. 
Let us consider the case when $K'=K+1$ and $I'=I$, $J'=J$.
This means that if the set $(\bar x_I, \bar y_J, \bar z_k) $ is represented by $N$
marked instants of time $\t_1< \t_2< \dots < \t_N$, then $\t_{N+1} > \t_n$ and $z_{K+1}= \t_{N+1}$. 
Obviously, the numbers $f''_{K+1} = f$ and $ \bar \vp^{(K+1)} = (\vp^{(K+1)}_1, \dots, \vp^{(K+1)}_{f}) $
are also joined to the set of parameters $\la \CP_R\ra$ (3.17).

Let us briefly describe the steps that we perform to get the estimate needed. 
Regarding the vertices and the edge-boxes of realization of the color diagram 
$\la \CG^{(c)}(\bar p, \bar q, \bar \nu)\ra_s$, 
we separate the edge-boxes of each vertex into two groups in dependence of whether the values in the boxes
are less than $\t_{N+1}$ or greater than $\t_{N+1}$. Clearly, the vertex attached by the edge-box with $\t_{N+1}$ plays a special role here.
By this procedure, we obtain realizations of two sub-diagrams $\la \grave \CG\ra$ and $\la \acute \CG\ra$ 
determined in obvious way. 

\vs 
The underlying trees $\CT_s =\CT(\CW_{2s})$ of the walks  are to be of the following structure: 
there exists a branch $\CB_{N+1}$ such that the descending path from the extreme vertex 
 $\check u_{N+1}$ to the root $\fb_0$ is of the total length 
not less  than $\vert \bar \phi^{(K+1)}\vert = \sum_{i=1}^{f} \vp^{(K+1)}_i$. At the vertex $\check \u_{N+1}$ 
and corresponding $f$ vertices of the descending part of $\CB_{N+1}$, the sub-clusters of the total number of $D_{N+1}$ edges
are attached. Then the remaining edges are used to construct sub-trees attached to $l_{N+1} + D_{N+1}- f$ vertices.
We denote this part of $\CT$ by $\acute \CT$ . 

\vs It is clear that the set of the walks under consideration can be represented in the form of the right-hand side of (5.12)
with $\t_2$ replaced by $\t_{N+1}$ and that the exponential estimate with the factor $e^{- \eta D_{N+1}}$ can be obtained
for the family of trees $\{\acute\CT\}$ (see also inequality (5.13), where $q''$ and $p''$ are determined with the help of 
sub-diagram $\acute\CG$). Using (5.17), it is not hard to 
complete the proof of (5.6) in the case of $N'=N+1$. We omit the detailed computations here because
they repeat in major part those performed earlier in this sub-section (see also \cite{K3} for more discussion of the general step
 of recurrent estimates).


\vskip 0.2cm {\bf Acknowledgements.} The author is grateful to V. Vengerovsky 
for the careful reading  of the manuscript and useful remarks.

\end{document}